\documentclass[aps,prl,twocolumn,amsmath,amssymb,floatfix,superscriptaddress,citeautoscript]{revtex4-2}

\usepackage{graphicx}
\usepackage{natbib}
\usepackage{hyperref}

\begin{document}

\title{Tuning the nontrivial topological properties of the Weyl semimetal CeAlSi}

\author{M. M. Piva}
\email{Mario.Piva@cpfs.mpg.de}
\affiliation{Max Planck Institute for Chemical Physics of Solids, N\"{o}thnitzer Str.\ 40, D-01187 Dresden, Germany}

\author{J. C. Souza}
\affiliation{Instituto de F\'{\i}sica ``Gleb Wataghin'', UNICAMP, 13083-859, Campinas, SP, Brazil}
\affiliation{Max Planck Institute for Chemical Physics of Solids, N\"{o}thnitzer Str.\ 40, D-01187 Dresden, Germany}

\author{V. Brousseau-Couture}
\affiliation{D\'{e}partement de Physique and Regroupement Qu\'{e}b\'{e}cois sur les Mat\'{e}riaux de Pointe, Universit\'{e} de Montr\'{e}al, Montr\'{e}al, Canada}

\author{K. R. Pakuszewski}
\affiliation{Instituto de F\'{\i}sica ``Gleb Wataghin'', UNICAMP, 13083-859, Campinas, SP, Brazil}

\author{Janas K. John}
\affiliation{Max Planck Institute for Chemical Physics of Solids, N\"{o}thnitzer Str.\ 40, D-01187 Dresden, Germany}

\author{C. Adriano}
\affiliation{Instituto de F\'{\i}sica ``Gleb Wataghin'', UNICAMP, 13083-859, Campinas, SP, Brazil}

\author{M. C\^{o}t\'{e}}
\affiliation{D\'{e}partement de Physique and Regroupement Qu\'{e}b\'{e}cois sur les Mat\'{e}riaux de Pointe, Universit\'{e} de Montr\'{e}al, Montr\'{e}al, Canada}

\author{P. G. Pagliuso}
\affiliation{Instituto de F\'{\i}sica ``Gleb Wataghin'', UNICAMP, 13083-859, Campinas, SP, Brazil}

\author{M. Nicklas}
\email{Michael.Nicklas@cpfs.mpg.de}
\affiliation{Max Planck Institute for Chemical Physics of Solids, N\"{o}thnitzer Str.\ 40, D-01187 Dresden, Germany}

\date{\today}

\begin{abstract}

In the ferromagnetic Weyl semimetal CeAlSi both space-inversion and time-reversal symmetries are broken. We use external pressure as an effective tuning parameter and relate three observations to the presence of a nontrivial topology in its ferromagnetic regime: an exceptional temperature response of the quantum oscillations amplitude, the presence of an anomalous Hall effect (AHE), and the existence of an unusual loop Hall effect (LHE). We find a suppression of the AHE and the LHE with increasing pressure, while the Curie temperature is enhanced. The magnetic structure and the electronic bands exhibit only a negligible pressure effect suggesting the importance of the domain wall landscape for the topological behavior in CeAlSi.

\end{abstract}

\maketitle

\newpage

%INTRODUCTION.

Topological phases of matter lately receive considerable attention, due to the experimental realization of novel types of charge carriers. One example are the massless  Weyl fermions found in Weyl semimetals (WSMs) \cite{yan2017topological,zhang2018towards,armitage2018weyl}. They are characterized by novel electronic properties, such as Fermi arcs, chiral anomaly, axial–gravitational anomaly, and an extremely large magnetoresistance (MR) \cite{yan2017topological,armitage2018weyl,yang2018symmetry,GoothNature2017}. Weyl fermions can be generated by breaking space-inversion (SI) or time-reversal (TR) symmetry of a Dirac material. So far most experimentally studied WSMs break the SI symmetry \cite{weng2015weyl,huang2015weyl,xu2015discovery,zhang2016signatures,huang2015observation,shekhar2015extremely,ali2014large,zhu2015quantum,jiang2017signature}. Fewer examples are known for Weyl semimetals with broken TR symmetry, i.e.\ \textit{magnetic} WSMs \cite{kuroda2017evidence,yang2017topological,morali2019fermi,liu2019magnetic,belopolski2019discovery,borisenko2019time}. They offer the potential to manipulate magnetically a protected topological phase in a desired way, e.g.\ for next-generation spintronics applications \cite{kurebayashi2016voltage,yang2021chiral}. They further provide new possibilities by combining topology and strong magnetic correlations \cite{grefe2020weyl,paschen2021quantum}.

The family of $Ln$Al$Pn$ ($Ln$ = lanthanides, $Pn$ = Ge, Si) materials is ideal to host nontrivial topological properties due to the noncentrosymmetric crystalline structure ($I4_{1}md$), the same as in the TaAs family of WSMs \cite{weng2015weyl,xu2015experimental,yang2015weyl,xu2015discovery,arnold2016negative,liu2016evolution}. In fact, multiple Weyl nodes and a large spin Hall effect were predicted to exist in LaAlGe and LaAlSi \cite{ng2021origin}. Weyl cones were experimentally observed for LaAlGe \cite{xu2017discovery} and a $\pi$~Berry phase was recently found in LaAlSi \cite{su2021multiple}. Furthermore, the magnetic members of the family also break TR symmetry in addition to SI symmetry and Weyl nodes were predicted to exist in many of the materials \cite{yang2020transition,xu2021shubnikov,chang2018magnetic,yang2021noncollinear}. Novel properties were also experimentally observed, such as an anomalous Hall effect (AHE) in PrAlGe$_{1-x}$Si$_{x}$ \cite{yang2020transition}, Fermi arcs in PrAlGe \cite{sanchez2020observation,destraz2020magnetism}, a topological magnetic phase and a singular angular MR in the semimetal CeAlGe \cite{hodovanets2018single,suzuki2019singular,puphal2020topological}, a Weyl mediated magnetism in NdAlSi \cite{gaudet2020incommensurate} and a $\pi$~Berry phase in SmAlSi \cite{xu2021shubnikov}.

CeAlSi is a member of the $Ln$Al$Pn$ family, which was earlier reported as crystallizing in the centrosymmetric $I4_{1}/amd$ structure \cite{flandorfer1998systems,pukas2004crystal,bobev2005ternary,sharma2007oscillations}. However, recent studies confirmed that the correct structure is the noncentrosymmetric $I4_{1}md$ \cite{yang2021noncollinear}, bringing new attention to this material. CeAlSi hosts an in-plane noncollinear ferromagnetic (FM) order below 8~K with a large anisotropy, the $c$-axis being the magnetically hard one. Moreover, several Weyl nodes are present close to the Fermi surface of CeAlSi, which give rise to an anisotropic Hall effect, yielding an AHE for field applied parallel to $[100]$ and a loop Hall effect (LHE) for field along the $[001]$ direction \cite{yang2021noncollinear}. 

An important property for the topological behaviors is the presence of a magnetoelastic coupling, in which the internal fields of the FM order generate picometer displacements in the unit cell  \cite{xu2021picoscale}. These lead to strains in the domain walls and create magnetic phases with different textures \cite{xu2021picoscale}. In fact, it is theoretically predicted that magnetic WSMs can host nontrivial domain walls \cite{huang2021topological} and chiral domain walls were indeed recently detected in CeAlSi \cite{sun2021mapping}. The presence of a magnetoelastic effect in CeAlSi suggests a strong response of AHE and LHE to the application of external pressure. Hydrostatic pressure has been also an effective tool in tuning the Fermi energy without introducing any additional disorder and was successfully used to tune the Weyl points closer to $E_{F}$ in many topological materials \cite{dos2016pressure,liang2017pressure,hirayama2015weyl,rodriguez2020two}. Moreover, application of pressure systematically  modifies the magnetic properties in Ce-based materials.

Here, we focus on the ferromagnetic Weyl semimetal CeAlSi. We use hydrostatic pressure as a tool to investigate the origin of the features characteristic of the nontrivial topological behavior in CeAlSi. For that we combine electrical transport and magnetization measurements with density-functional theory (DFT) calculations. Our results reveal three features possibly associated with the presence of a nontrivial Berry phase: an unexpected temperature response of the quantum oscillations (QO) amplitude, a large anomalous Hall effect and the presence of a loop Hall effect, which are all suppressed by application of hydrostatic pressure. While we find a linear increase of the Curie temperature ($T_C$) upon application of pressure, our magnetic measurements and theoretical calculations indicate the absence of changes in the magnetic structure and only negligible changes in the electronic band structure. Our results therefore support the relevance of magnetic domain walls for the presence of the topological behavior in CeAlSi and we conclude that external pressure favors trivial domain walls over nontrivial ones.

 \begin{figure}[!t]
	\includegraphics[width=0.85\linewidth]{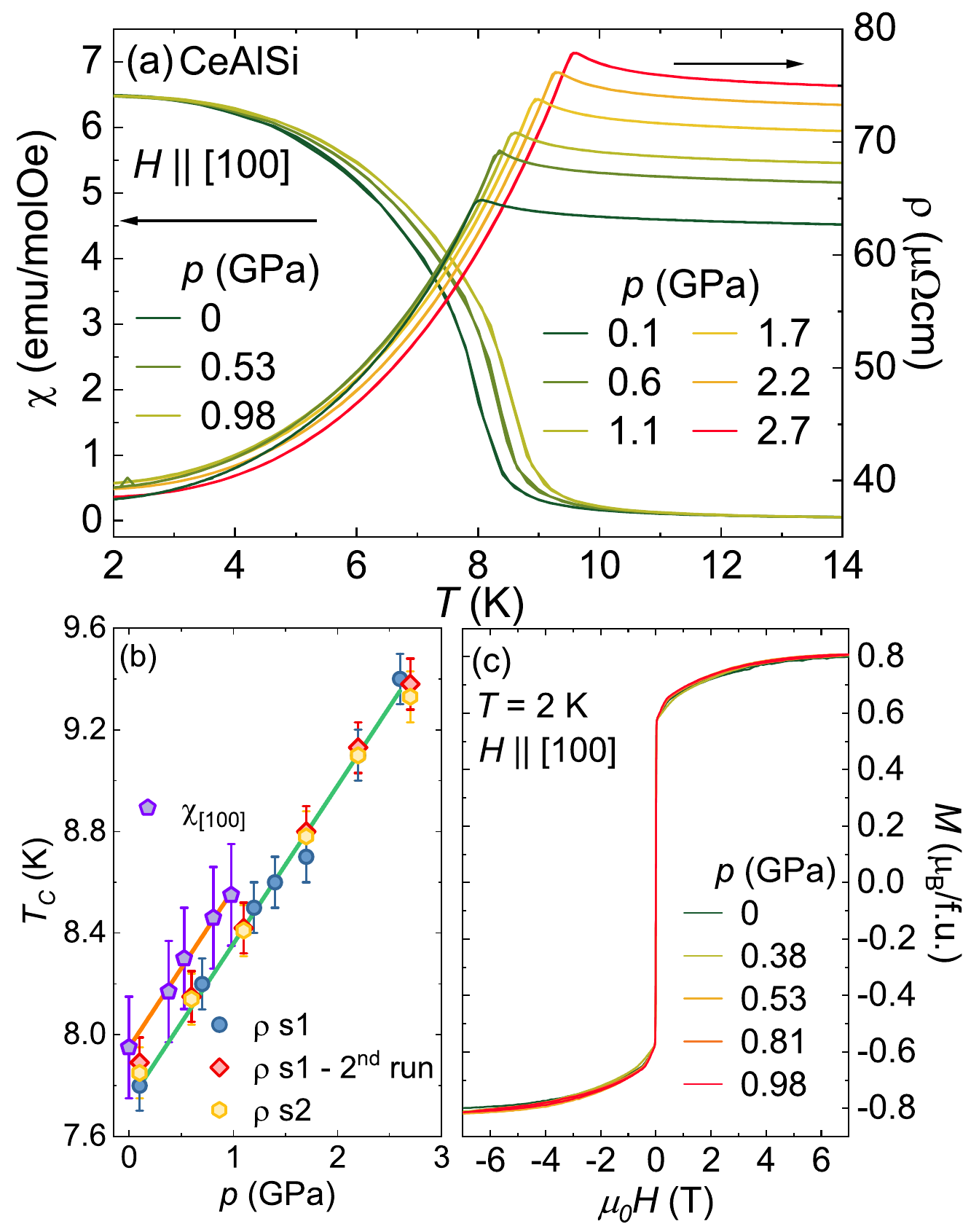}
	\caption{(a)  Magnetic susceptibly $\chi=M/H$ (left axis), obtained in an applied field of 50~mT along the $[100]$ crystal axis, and electrical resistivity (right axis) as a function of temperature for selected pressures. %Note that $1$~emu $=$ $10^{-3}$~Am$^{2}$.
 (b) Temperature--pressure phase diagram. The solid lines are linear fits. (c) Magnetization measurements for several pressures.}
	\label{chi_rho}
\end{figure}

\begin{figure}[!t]
	\includegraphics[width=0.85\linewidth]{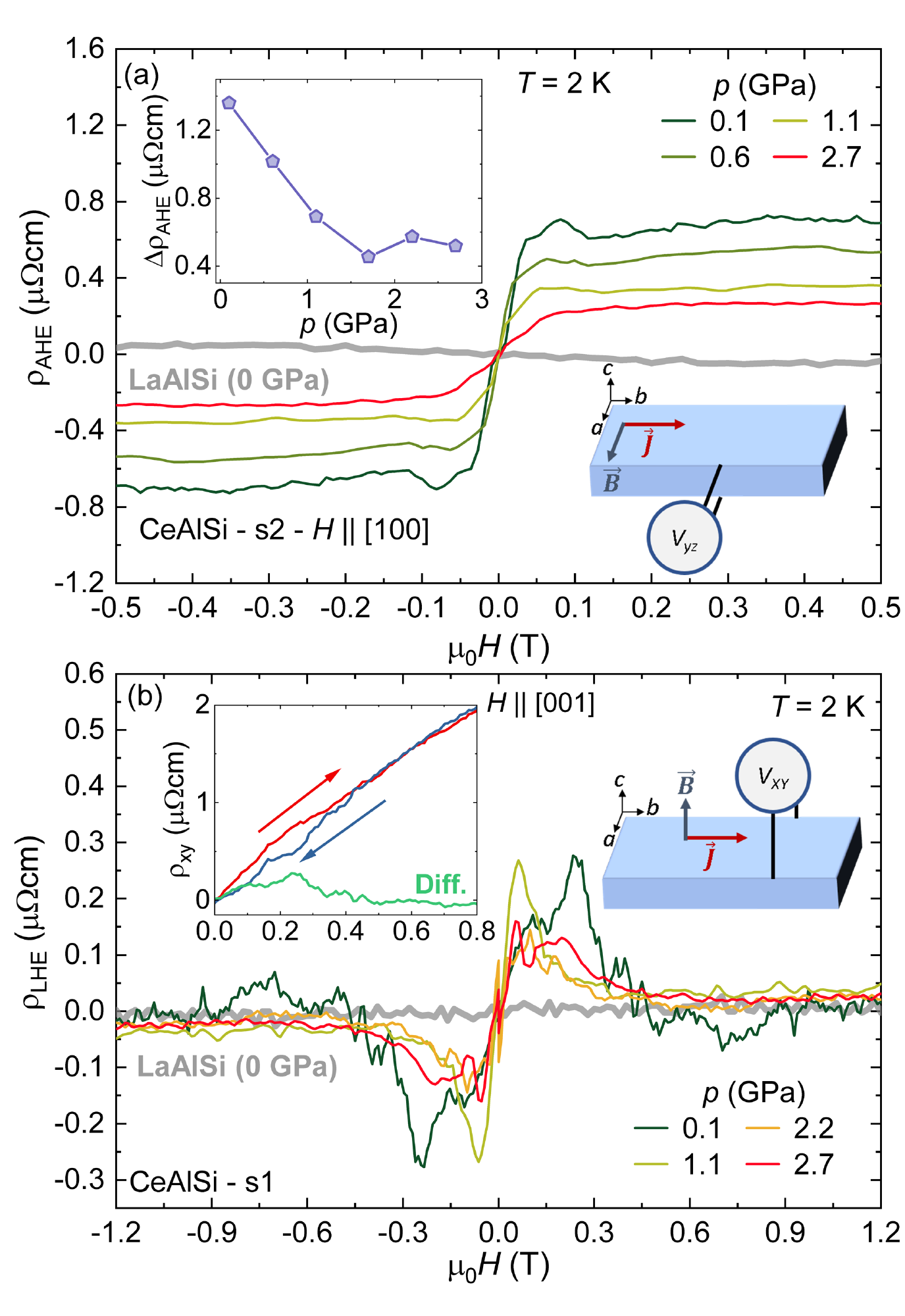}
	\caption{ (a) Anomalous Hall effect (AHE) at 2~K as a function of magnetic fields $H\parallel[100]$ for different applied pressures. The top inset shows the AHE jump as a function of pressure and the bottom inset displays a scheme of the circuit used in this measurement. s1 and s2 denote samples 1 and 2, respectively.
(b) Loop Hall effect (LHE) at 2~K as a function of magnetic field for $H\parallel[001]$ for selected applied pressures. The left inset shows the Hall resistivity measured upon increasing (red) and decreasing (blue) magnetic field and the difference of both curves (green) at 0.1~GPa. The right inset displays a schematic drawing of the measurement circuit. Data of the nonmagnetic reference material LaSiAl at ambient pressure is shown as gray line in both panels}
	\label{LHE_AHE}
\end{figure}

%RESULTS.

%Pressure characterization

At ambient pressure, magnetic susceptibility $\chi(T)$ and electrical resistivity $\rho(T)$ data taken on different CeAlSi samples indicate FM ordering below $T_{C} \approx 8$~K [see Fig.~\ref{chi_rho}(a)], in good agreement with previous reports \cite{yang2021noncollinear,xu2021picoscale,sun2021mapping}. The application of external pressure linearly enhances $T_C(p)$ with a slope of 0.62(2)~K/GPa, driving $T_{C}$ from 7.8~K at ambient pressure to 9.4~K at 2.7~GPa (values taken from the resistivity data), as shown in Fig.~\ref{chi_rho}(b). We note that $\rho(T)$ displays a metallic behavior in the full pressure range at temperatures up to 300~K. At high temperatures, a broad hump in $\rho(T)$ around 80~K can be associated with scattering originating from the thermal population of the first excited crystal-electric field (CEF) level. The position of the hump is almost unchanged as function of pressure, indicating that the position of the first excited CEF state does not change with pressure. More important is our finding that for different pressures the in-plane magnetization curves $M(H)$ at 2~K with magnetic field along the $[100]$ direction lie on top of each other [see Fig.\ \ref{chi_rho}(c)], indicating that pressure has a negligible effect on the noncollinear planar magnetic structure found at ambient pressure \cite{yang2021noncollinear}. With this background we turn to the pressure dependence of the topological features observed in CeAlSi.

%Hall effect

We first focus on the  anomalous Hall effect (AHE) shown in Fig.~\ref{LHE_AHE}(a). $\rho_{\rm AHE}(H)$ was obtained by subtracting a linear background from the measured Hall resistivity $\rho_{yz}$ considering $\rho_{yz} = R_{0}H + \rho_{\rm AHE}$. Here $R_{0}$ is the ordinary contribution to the Hall effect. $R_{0}$ was determined by a linear fit to $\rho_{yz}(H)$ in the range $0.2~{\rm T}\leqslant \mu_0H \leqslant 0.6~{\rm T}$. The evolution of $R_{0}$ as a function of pressure yields a decrease in the electron density from $6.0 \times 10^{20}$~cm$^{-3}$ at ambient pressure to $1.1 \times 10^{20}~{\rm cm^{-3}}$ at 2.7~GPa. At ambient pressure, a large AHE is observed in the ferromagnetic regime of CeAlSi. It is absent in the non-magnetic analog LaAlSi and in the paramagnetic region of CeAlSi. Application of external pressure suppresses the size of the jump in the AHE, which is the difference in $\rho_{\rm AHE}$ between positive and negative magnetic fields [top inset of Fig.~\ref{LHE_AHE}(a)]. This reduction of the AHE upon increasing pressure is a strong indication of its topological origin. As we have shown above, even though $T_C(p)$ increases, the $M(H)$ curves remain unchanged upon increasing pressure [see Fig.~\ref{chi_rho}(c)]. This rules out ferromagnetism as the source of the AHE signal, in favor of modifications in the nontrivial topological features of CeAlSi.

The second feature in CeAlSi, which is believed to be caused by the nontrivial topology is the loop Hall effect displayed in Fig.~\ref{LHE_AHE}(b). In CeAlSi, this unusual effect is seen for magnetic fields applied in parallel to the $[001]$ direction. It is important to remark that it is only observed in the ferromagnetic regime and also absent in the non-magnetic analog LaAlSi. $\rho_{\rm LHE}$ is obtained by recording the Hall resistivity $\rho_{xy}$ for $H\parallel[001]$ upon increasing and decreasing magnetic field and taking the difference between both curves, as shown in the left inset of Fig.~\ref{LHE_AHE}(b) for 0.1~GPa.
We note that the magnitude of the LHE in our data at ambient pressure is comparable to that one previously reported \cite{yang2021noncollinear}. Its existence in CeSiAl can be phenomenologically  related to the presence of the Weyl nodes near the Fermi energy suggesting its connection to the nontrivial topology in CeAlSi \cite{yang2021noncollinear}.
Application of external pressure suppresses the LHE [see Fig.~\ref{LHE_AHE}(b)]. Due to the smallness of the signal and the corresponding noise level, it is not possible to quantitatively evaluate the size of the suppression.

An LHE was also observed in the pyrochlore iridate Nd$_{2}$Ir$_{2}$O$_{7}$ \cite{ma2015mobile,ueda2018spontaneous}. Nd$_{2}$Ir$_{2}$O$_{7}$ is an insulating material close to a WSM phase, in which the LHE was only observed in a narrow temperature range close to the antiferromagnetic transition temperature, where all-in-all-out and all-out-all-in domains were present in the material \cite{ma2015mobile,ueda2018spontaneous}. The all-in-all-out domains were predicted to host Weyl fermions and surface states, which are suppressed at very low temperatures making the material insulating. However, mid-gap states survive in the magnetic domain walls \cite{yamaji2014metallic}, possibly leading to the highly conductive domain walls and the presence of the LHE in Nd$_{2}$Ir$_{2}$O$_{7}$ \cite{ma2015mobile,ueda2018spontaneous}. A similar scenario might occur in CeAlSi, where recently chiral domain walls were found and might be related to the observed topological properties \cite{sun2021mapping}. Furthermore, nontrivial domain walls are predicted to exist in magnetic WSMs \cite{huang2021topological}. It is conceivable that the internal fields already create tiny displacements at ambient pressure leading to differently oriented magnetic regions due to the presence of strain at domain walls via magnetostriction/magnetoelastic effects \cite{xu2021picoscale}. A compression of the sample by the application of hydrostatic pressure could then effectively change the landscape of the magnetic domain walls in CeAlSi, modifying its topological properties.

% Quantum oscillations

\begin{figure}[!t]
	\includegraphics[width=0.85\linewidth]{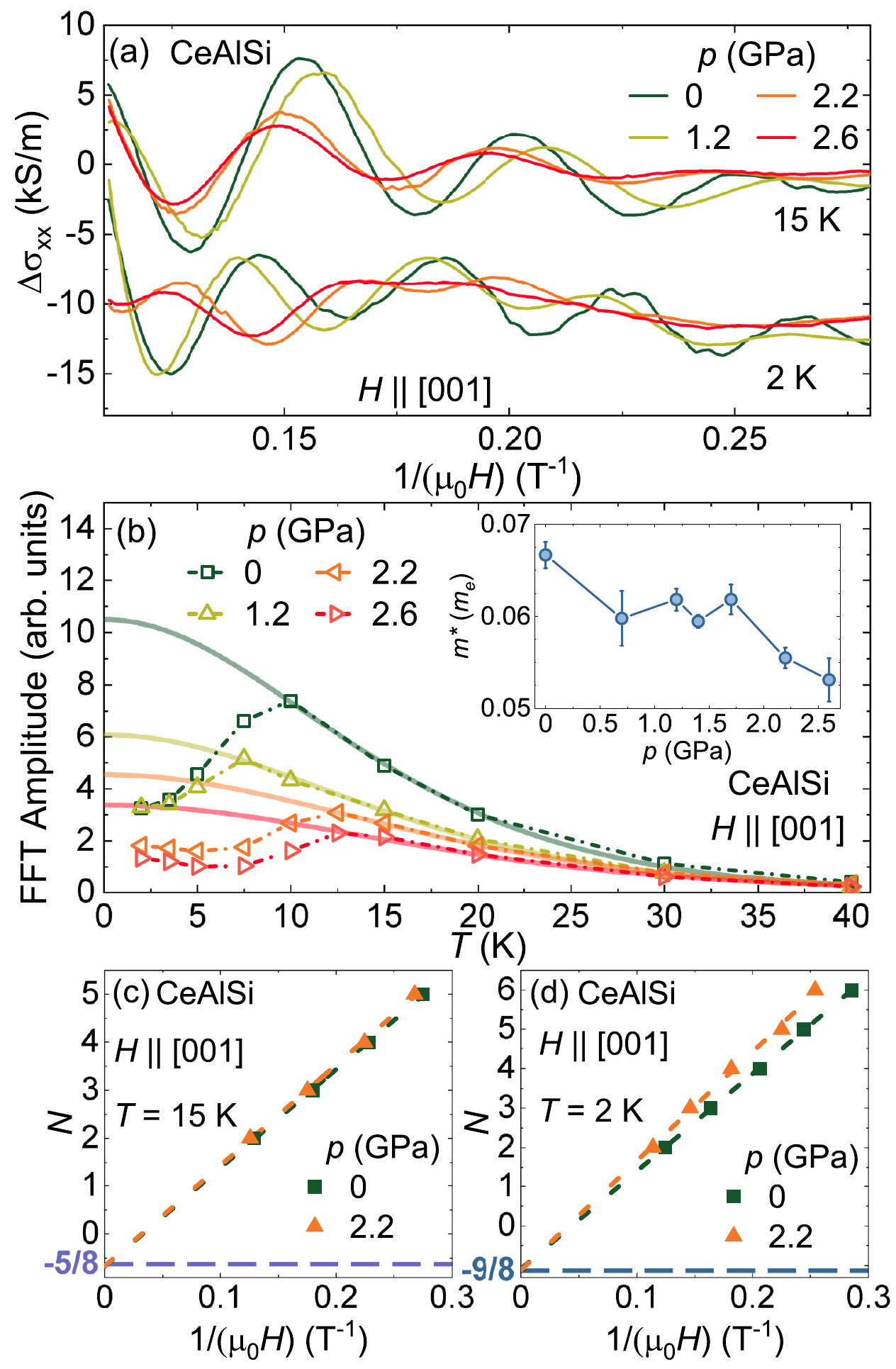}
	\caption{(a) Longitudinal conductivity $H\parallel[001]$ after subtraction of a third order polynomial background $\Delta\sigma_{xx}$ as a function of $1/(\mu_{0}H)$ at 15~K (top) and 2~K (bottom) for selected pressures. The curves at 2~K were shifted by $-10$~kS/m for clarity. (b) Fast Fourier transformation (FFT) amplitude as a function of temperature at different applied pressures. The solid lines are simulations considering the best fits using the Lifshitz-Kosevich formula. The inset shows the effective mass $m^{*}$ as a function of pressure. (c) and (d) Landau fan diagrams for CeAlSi at 15 and 2~K, respectively.}
	\label{DeltaSigma}
\end{figure}

Finally, we found an unexpected behavior of the quantum oscillations in CeAlSi upon cooling, which has not been reported before. We argue that it can be taken as an indication for the presence of a nontrivial $\pi$~Berry phase. Figure~\ref{DeltaSigma}(a) displays the QO well above the Curie temperature at 15~K and deep inside the ferromagnetically ordered state at 2~K at several pressures. $\Delta\sigma_{xx}$ was obtained by subtracting a third-order polynomial background from the conductivity, which was extracted by inverting the resistivity tensor for tetragonal structures $\sigma_{xx} = \rho_{xx}/(\rho_{xx}^{2} + \rho_{xy}^{2})$ \cite{lovett2018tensor}. The oscillations are clearly visible up to 40~K at all studied pressures.
We notice two main features: i) the amplitudes of the oscillations at 15~K are larger than those at 2~K and ii) the amplitude of the oscillations is suppressed by increasing pressure.

A fast Fourier transformation (FFT) of $\Delta\sigma_{xx}$ in the available field range reveals only one frequency $f\approx 20(5)$~T. Increasing pressure suppresses the FFT amplitude [Fig.~\ref{DeltaSigma}(b)]. The reduction in the FFT amplitude in the ferromagnetically ordered state, already visible in the raw data, becomes clearly evident in Fig.~\ref{DeltaSigma}(b). We note that the frequency does not change upon entering the ferromagnetic phase.
Such a decline upon decreasing temperature is very unusual. Generally, the thermal damping of the FFT amplitude can be described by the Lifshitz-Kosevich (LK) formula \cite{shoenberg2009magnetic}
\begin{equation}
R_{T} = \frac{\alpha T m^{*}}{B \sinh (\alpha T m^{*}/B)},
\end{equation}
in which $\alpha=2 \pi^{2} k_{B}/e \hbar \approx 14.69$~T/K, $T$ is the temperature, $B$ is the magnetic and $m^{*}$ the effective mass. In the paramagnetic regime, the amplitude at different pressures follows the predicted LK behavior [solid lines in Fig.~\ref{DeltaSigma}(b)].
The effective mass $m^{*}$ was extracted in the paramagnetic region using the LK formula and is presented in the inset of Fig.~\ref{DeltaSigma}(b).

Once ferromagnetism starts to develop upon cooling, the temperature dependence of the amplitude begins to deviate from the behavior described by the LK formula. That is observed for all studied pressures. This rare response of the oscillation amplitude as a function of temperature was not observed in other members of the $Ln$Al$Pn$ family so far \cite{yang2021noncollinear,su2021multiple,xu2021shubnikov,gaudet2020incommensurate}. It was previously observed in just a few materials, such as the charge-transfer salts \cite{honold1997importance} and SmSb \cite{wu2019anomalous}. In the first case, the sudden decrease of the amplitude with decreasing temperature was associated with the presence of highly metallic edge states related to the quantum Hall effect in quasi-two-dimensional materials \cite{honold1997importance}, which is unlikely to be the cause for the behavior found in CeAlSi.
In the latter system, a sudden decrease of the Shubnikov-de Haas oscillations takes place once the material becomes antiferromagnetic, which could be caused by a nontrivial Berry phase \cite{wu2019anomalous}. Similarly to CeAlSi, the unusual behavior of the QO amplitude also appears in connection with a magnetically ordered phase.

The further analysis of the QO data indicates that the appearance of ferromagnetism in CeAlSi induces a change in the nature of the topological properties, as indicated by the Landau fan analysis shown in Figs.~\ref{DeltaSigma}(c) and \ref{DeltaSigma}(d). At 15~K, CeAlSi  presents an intercept around $-5/8$, which indicates the presence of trivial charge carriers \cite{wang2016anomalous}. However, at 2~K, the intercept is at $-9/8$, which for 3D magnetic WSMs can be associated with linear dispersive charge carriers and a nontrivial $\pi$~Berry phase \cite{wang2016anomalous}. Moreover, the intercept at $-9/8$ maybe also an indication that the Weyl cones are close to, but not exactly at $E_{F}$, which would yield a $-1/8$ intercept. This finding agrees with the above discussed nontrivial topological features found in the presence of the AHE and LHE. It is also consistent with the fact that these features are not completely suppressed in the studied pressure range, as the $-9/8$ intercept at 2~K persists up to 2.2~GPa. Above that pressure, the QO become too weak for a reliable Landau fan analysis.

%Band structure

\begin{figure}
	\centering
	\includegraphics[width=0.95\linewidth]{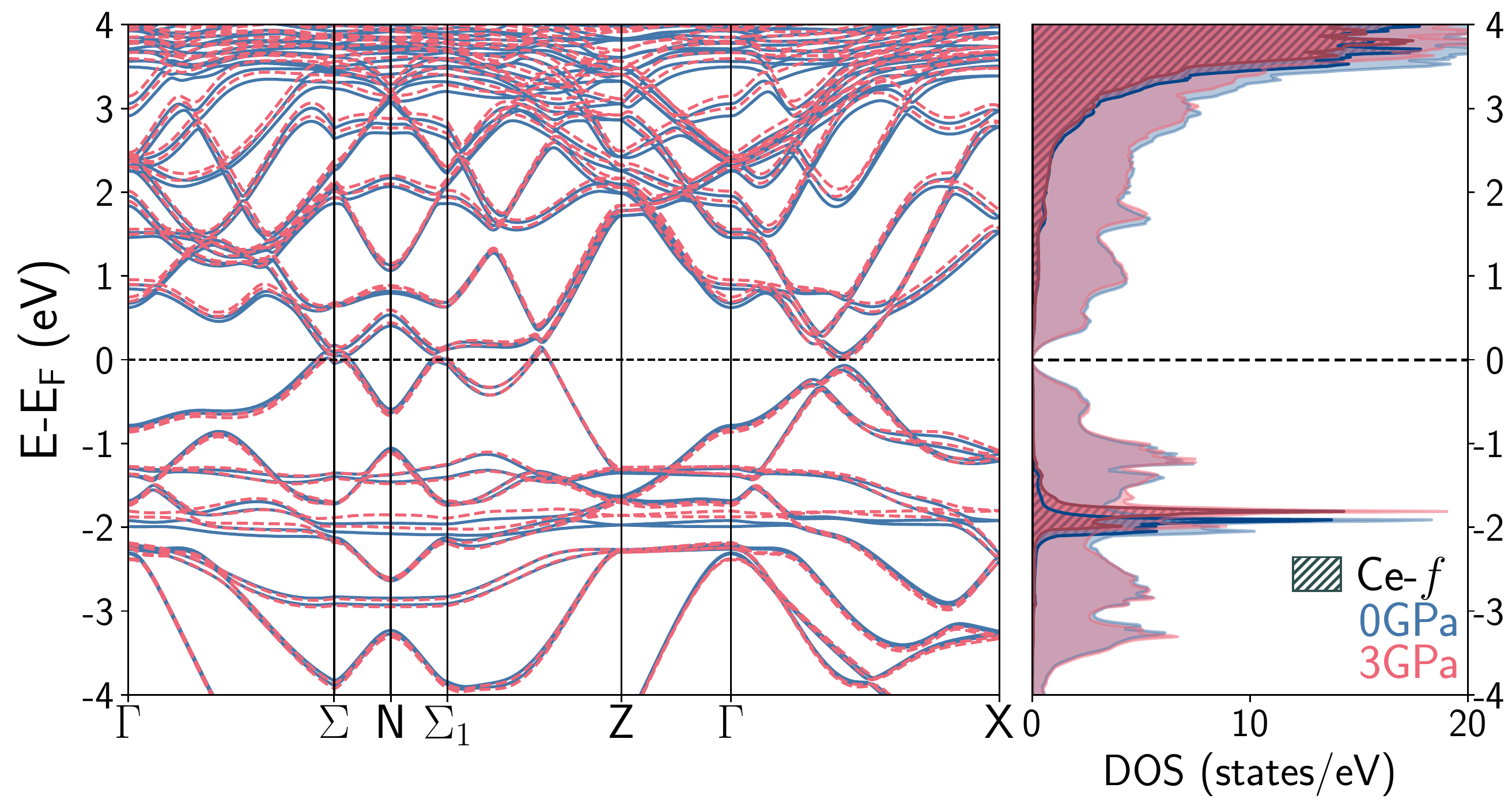}
	\caption{Electronic bands and DOS at ambient pressure (blue) and 3~GPa (red). The hatched region of right panel corresponds to the partial DOS associated with Ce $f$ states.}
	\label{fig:bstr_with_f}
\end{figure}

So far we did not consider possible modifications of the electronic band structure as a function of pressure. These could provide a further way to modify the topological properties in CeAlSi: the Weyl nodes present near the Fermi surface may be further pushed away from the Fermi level upon increasing pressure. This would also lead to a suppression of the observed AHE and LHE, but hardly affect the intercept observed in the Landau fan diagrams. DFT calculations at 0 and 3~GPa indicate only a negligible effect of pressure on the electronic bands and the electronic density of states (DOS) (see Fig.\ \ref{fig:bstr_with_f}). The bands contributing to the hole pockets on the Fermi surface barely display any variation of their intercepts of the Fermi energy in $\bm{k}$-space, suggesting a negligible variation of the Fermi surface area. The only noticeable modification of the electronic structure throughout the investigated pressure range
is a small shift of the bands associated with Ce $f$-electrons to higher energies with respect to the Fermi energy. As these bands lie about 2~eV below the Fermi level, they most likely do not contribute to the AHE and LHE. Therefore, this excludes pressure-induced modifications of the band structure as source of the observed suppression of the LHE and AHE. Thus, the band structure calculations support the importance of the magnetic domain walls and their sensitivity to external pressure for the nontrivial topological properties observed in CeAlSi.

%CONCLUSIONS.

In summary, our results revealed three different properties associated with the possible presence of a nontrivial Berry phase in the Weyl semimetal CeAlSi and their pressure dependence: i) an exceptional temperature dependence of the QO amplitude in the ferromagnetic regime and ii) an anomalous Hall effect and ii) a loop Hall effect. All of them are closely connected to the ferromagnetically ordered phase present in CeAlSi, as none of them is observed in the non-magnetic analog LaAlSi and at temperatures above $T_{C}$. This directly indicates the fundamental role of the ferromagnetism and the connected breaking of the TR symmetry in addition to the broken SI symmetry in the WSM CeAlSi. Application of external pressure suppresses the magnitude of the anomalous Hall effect, loop Hall effect, and also the magnitude of the quantum oscillations. These observations suggest that compressing CeAlSi changes its magnetic domain landscape via magnetoelastic couplings, which lead to magnetostriction effects favoring the trivial domain walls over the nontrivial ones. Our findings in CeAlSi propose that application of tensile strain as a prospective route for enhancing the nontrivial topological features of CeAlSi and related materials.

\begin{acknowledgments}
We thank U.\ Burkhardt for carrying out energy dispersive x-ray analysis on the samples. This work was supported by the S\~ao Paulo Research Foundation (FAPESP) grants 2017/10581-1, 2018/11364-7, 2020/12283-0, CNPq grants $\#$ 304496/2017-0, 310373/2019-0 and CAPES, Brazil. This research was financially supported by the Natural Sciences and Engineering Research Council of Canada (NSERC), under the Discovery Grants program grant No. RGPIN-2016-06666. Computations were made on the supercomputer Beluga managed by Calcul Qu\'ebec and Compute Canada. The operation of these supercomputers is funded by the
Canada Foundation for Innovation, the Minist\`ere de la Science, de l'\'Economie et de l'Innovation du Qu\'ebec, and the Fonds de recherche du Qu\'ebec – Nature et technologies.
\end{acknowledgments}

\bibliography{CeAlSi}

\begin{thebibliography}{60}%
\makeatletter
\providecommand \@ifxundefined [1]{%
 \@ifx{#1\undefined}
}%
\providecommand \@ifnum [1]{%
 \ifnum #1\expandafter \@firstoftwo
 \else \expandafter \@secondoftwo
 \fi
}%
\providecommand \@ifx [1]{%
 \ifx #1\expandafter \@firstoftwo
 \else \expandafter \@secondoftwo
 \fi
}%
\providecommand \natexlab [1]{#1}%
\providecommand \enquote  [1]{``#1''}%
\providecommand \bibnamefont  [1]{#1}%
\providecommand \bibfnamefont [1]{#1}%
\providecommand \citenamefont [1]{#1}%
\providecommand \href@noop [0]{\@secondoftwo}%
\providecommand \href [0]{\begingroup \@sanitize@url \@href}%
\providecommand \@href[1]{\@@startlink{#1}\@@href}%
\providecommand \@@href[1]{\endgroup#1\@@endlink}%
\providecommand \@sanitize@url [0]{\catcode `\\12\catcode `\$12\catcode
  `\&12\catcode `\#12\catcode `\^12\catcode `\_12\catcode `\%12\relax}%
\providecommand \@@startlink[1]{}%
\providecommand \@@endlink[0]{}%
\providecommand \url  [0]{\begingroup\@sanitize@url \@url }%
\providecommand \@url [1]{\endgroup\@href {#1}{\urlprefix }}%
\providecommand \urlprefix  [0]{URL }%
\providecommand \Eprint [0]{\href }%
\providecommand \doibase [0]{https://doi.org/}%
\providecommand \selectlanguage [0]{\@gobble}%
\providecommand \bibinfo  [0]{\@secondoftwo}%
\providecommand \bibfield  [0]{\@secondoftwo}%
\providecommand \translation [1]{[#1]}%
\providecommand \BibitemOpen [0]{}%
\providecommand \bibitemStop [0]{}%
\providecommand \bibitemNoStop [0]{.\EOS\space}%
\providecommand \EOS [0]{\spacefactor3000\relax}%
\providecommand \BibitemShut  [1]{\csname bibitem#1\endcsname}%
\let\auto@bib@innerbib\@empty
%</preamble>
\bibitem [{\citenamefont {Yan}\ and\ \citenamefont
  {Felser}(2017)}]{yan2017topological}%
  \BibitemOpen
  \bibfield  {author} {\bibinfo {author} {\bibfnamefont {B.}~\bibnamefont
  {Yan}}\ and\ \bibinfo {author} {\bibfnamefont {C.}~\bibnamefont {Felser}},\
  }\bibfield  {title} {\bibinfo {title} {{Topological Materials: Weyl
  Semimetals}},\ }\href
  {https://doi.org/10.1146/annurev-conmatphys-031016-025458} {\bibfield
  {journal} {\bibinfo  {journal} {Annu. Rev. Condens. Matter Phys.}\ }\textbf
  {\bibinfo {volume} {8}},\ \bibinfo {pages} {337} (\bibinfo {year}
  {2017})}\BibitemShut {NoStop}%
\bibitem [{\citenamefont {Zhang}\ \emph {et~al.}(2018)\citenamefont {Zhang},
  \citenamefont {Lu}, \citenamefont {Shen}, \citenamefont {Chen},\ and\
  \citenamefont {Xiu}}]{zhang2018towards}%
  \BibitemOpen
  \bibfield  {author} {\bibinfo {author} {\bibfnamefont {C.}~\bibnamefont
  {Zhang}}, \bibinfo {author} {\bibfnamefont {H.-Z.}\ \bibnamefont {Lu}},
  \bibinfo {author} {\bibfnamefont {S.-Q.}\ \bibnamefont {Shen}}, \bibinfo
  {author} {\bibfnamefont {Y.~P.}\ \bibnamefont {Chen}},\ and\ \bibinfo
  {author} {\bibfnamefont {F.}~\bibnamefont {Xiu}},\ }\bibfield  {title}
  {\bibinfo {title} {{Towards the manipulation of topological states of matter:
  a perspective from electron transport}},\ }\href
  {https://doi.org/https://doi.org/10.1016/j.scib.2018.04.007} {\bibfield
  {journal} {\bibinfo  {journal} {Sci. Bull.}\ }\textbf {\bibinfo {volume}
  {63}},\ \bibinfo {pages} {580} (\bibinfo {year} {2018})}\BibitemShut
  {NoStop}%
\bibitem [{\citenamefont {Armitage}\ \emph {et~al.}(2018)\citenamefont
  {Armitage}, \citenamefont {Mele},\ and\ \citenamefont
  {Vishwanath}}]{armitage2018weyl}%
  \BibitemOpen
  \bibfield  {author} {\bibinfo {author} {\bibfnamefont {N.~P.}\ \bibnamefont
  {Armitage}}, \bibinfo {author} {\bibfnamefont {E.~J.}\ \bibnamefont {Mele}},\
  and\ \bibinfo {author} {\bibfnamefont {A.}~\bibnamefont {Vishwanath}},\
  }\bibfield  {title} {\bibinfo {title} {{Weyl and Dirac semimetals in
  three-dimensional solids}},\ }\href
  {https://doi.org/10.1103/RevModPhys.90.015001} {\bibfield  {journal}
  {\bibinfo  {journal} {Rev. Mod. Phys.}\ }\textbf {\bibinfo {volume} {90}},\
  \bibinfo {pages} {015001} (\bibinfo {year} {2018})}\BibitemShut {NoStop}%
\bibitem [{\citenamefont {Yang}\ \emph {et~al.}(2018)\citenamefont {Yang},
  \citenamefont {Yang}, \citenamefont {Derunova}, \citenamefont {Parkin},
  \citenamefont {Yan},\ and\ \citenamefont {Ali}}]{yang2018symmetry}%
  \BibitemOpen
  \bibfield  {author} {\bibinfo {author} {\bibfnamefont {S.-Y.}\ \bibnamefont
  {Yang}}, \bibinfo {author} {\bibfnamefont {H.}~\bibnamefont {Yang}}, \bibinfo
  {author} {\bibfnamefont {E.}~\bibnamefont {Derunova}}, \bibinfo {author}
  {\bibfnamefont {S.~S.~P.}\ \bibnamefont {Parkin}}, \bibinfo {author}
  {\bibfnamefont {B.}~\bibnamefont {Yan}},\ and\ \bibinfo {author}
  {\bibfnamefont {M.~N.}\ \bibnamefont {Ali}},\ }\bibfield  {title} {\bibinfo
  {title} {{Symmetry demanded topological nodal-line materials}},\ }\href
  {https://doi.org/10.1080/23746149.2017.1414631} {\bibfield  {journal}
  {\bibinfo  {journal} {Adv. Phys. X}\ }\textbf {\bibinfo {volume} {3}},\
  \bibinfo {pages} {1414631} (\bibinfo {year} {2018})}\BibitemShut {NoStop}%
\bibitem [{\citenamefont {Gooth}\ \emph {et~al.}(2017)\citenamefont {Gooth},
  \citenamefont {Niemann}, \citenamefont {Meng}, \citenamefont {Grushin},
  \citenamefont {Landsteiner}, \citenamefont {Gotsmann}, \citenamefont
  {Menges}, \citenamefont {Schmidt}, \citenamefont {Shekhar}, \citenamefont
  {Suess}, \citenamefont {Huehne}, \citenamefont {Rellinghaus}, \citenamefont
  {Felser}, \citenamefont {Yan},\ and\ \citenamefont
  {Nielsch}}]{GoothNature2017}%
  \BibitemOpen
  \bibfield  {author} {\bibinfo {author} {\bibfnamefont {J.}~\bibnamefont
  {Gooth}}, \bibinfo {author} {\bibfnamefont {A.~C.}\ \bibnamefont {Niemann}},
  \bibinfo {author} {\bibfnamefont {T.}~\bibnamefont {Meng}}, \bibinfo {author}
  {\bibfnamefont {A.~G.}\ \bibnamefont {Grushin}}, \bibinfo {author}
  {\bibfnamefont {K.}~\bibnamefont {Landsteiner}}, \bibinfo {author}
  {\bibfnamefont {B.}~\bibnamefont {Gotsmann}}, \bibinfo {author}
  {\bibfnamefont {F.}~\bibnamefont {Menges}}, \bibinfo {author} {\bibfnamefont
  {M.}~\bibnamefont {Schmidt}}, \bibinfo {author} {\bibfnamefont
  {C.}~\bibnamefont {Shekhar}}, \bibinfo {author} {\bibfnamefont
  {V.}~\bibnamefont {Suess}}, \bibinfo {author} {\bibfnamefont
  {R.}~\bibnamefont {Huehne}}, \bibinfo {author} {\bibfnamefont
  {B.}~\bibnamefont {Rellinghaus}}, \bibinfo {author} {\bibfnamefont
  {C.}~\bibnamefont {Felser}}, \bibinfo {author} {\bibfnamefont
  {B.}~\bibnamefont {Yan}},\ and\ \bibinfo {author} {\bibfnamefont
  {K.}~\bibnamefont {Nielsch}},\ }\bibfield  {title} {\bibinfo {title}
  {{Experimental signatures of the mixed axial-gravitational anomaly in the
  Weyl semimetal NbP}},\ }\href {https://doi.org/10.1038/nature23005}
  {\bibfield  {journal} {\bibinfo  {journal} {Nature}\ }\textbf {\bibinfo
  {volume} {547}},\ \bibinfo {pages} {324} (\bibinfo {year}
  {2017})}\BibitemShut {NoStop}%
\bibitem [{\citenamefont {Weng}\ \emph {et~al.}(2015)\citenamefont {Weng},
  \citenamefont {Fang}, \citenamefont {Fang}, \citenamefont {Bernevig},\ and\
  \citenamefont {Dai}}]{weng2015weyl}%
  \BibitemOpen
  \bibfield  {author} {\bibinfo {author} {\bibfnamefont {H.}~\bibnamefont
  {Weng}}, \bibinfo {author} {\bibfnamefont {C.}~\bibnamefont {Fang}}, \bibinfo
  {author} {\bibfnamefont {Z.}~\bibnamefont {Fang}}, \bibinfo {author}
  {\bibfnamefont {B.~A.}\ \bibnamefont {Bernevig}},\ and\ \bibinfo {author}
  {\bibfnamefont {X.}~\bibnamefont {Dai}},\ }\bibfield  {title} {\bibinfo
  {title} {{Weyl Semimetal Phase in Noncentrosymmetric Transition-Metal
  Monophosphides}},\ }\href {https://doi.org/10.1103/PhysRevX.5.011029}
  {\bibfield  {journal} {\bibinfo  {journal} {Phys. Rev. X}\ }\textbf {\bibinfo
  {volume} {5}},\ \bibinfo {pages} {011029} (\bibinfo {year}
  {2015})}\BibitemShut {NoStop}%
\bibitem [{\citenamefont {Huang}\ \emph
  {et~al.}(2015{\natexlab{a}})\citenamefont {Huang}, \citenamefont {Xu},
  \citenamefont {Belopolski}, \citenamefont {Lee}, \citenamefont {Chang},
  \citenamefont {Wang}, \citenamefont {Alidoust}, \citenamefont {Bian},
  \citenamefont {Neupane}, \citenamefont {Zhang} \emph
  {et~al.}}]{huang2015weyl}%
  \BibitemOpen
  \bibfield  {author} {\bibinfo {author} {\bibfnamefont {S.-M.}\ \bibnamefont
  {Huang}}, \bibinfo {author} {\bibfnamefont {S.-Y.}\ \bibnamefont {Xu}},
  \bibinfo {author} {\bibfnamefont {I.}~\bibnamefont {Belopolski}}, \bibinfo
  {author} {\bibfnamefont {C.-C.}\ \bibnamefont {Lee}}, \bibinfo {author}
  {\bibfnamefont {G.}~\bibnamefont {Chang}}, \bibinfo {author} {\bibfnamefont
  {B.}~\bibnamefont {Wang}}, \bibinfo {author} {\bibfnamefont {N.}~\bibnamefont
  {Alidoust}}, \bibinfo {author} {\bibfnamefont {G.}~\bibnamefont {Bian}},
  \bibinfo {author} {\bibfnamefont {M.}~\bibnamefont {Neupane}}, \bibinfo
  {author} {\bibfnamefont {C.}~\bibnamefont {Zhang}}, \emph {et~al.},\
  }\bibfield  {title} {\bibinfo {title} {{A Weyl Fermion semimetal with surface
  Fermi arcs in the transition metal monopnictide TaAs class}},\ }\href
  {https://doi.org/10.1038/ncomms8373} {\bibfield  {journal} {\bibinfo
  {journal} {Nat. Commun.}\ }\textbf {\bibinfo {volume} {6}},\ \bibinfo {pages}
  {1} (\bibinfo {year} {2015}{\natexlab{a}})}\BibitemShut {NoStop}%
\bibitem [{\citenamefont {Xu}\ \emph {et~al.}(2015{\natexlab{a}})\citenamefont
  {Xu}, \citenamefont {Belopolski}, \citenamefont {Alidoust}, \citenamefont
  {Neupane}, \citenamefont {Bian}, \citenamefont {Zhang}, \citenamefont
  {Sankar}, \citenamefont {Chang}, \citenamefont {Yuan}, \citenamefont {Lee},
  \citenamefont {Huang}, \citenamefont {Zheng}, \citenamefont {Ma},
  \citenamefont {Sanchez}, \citenamefont {Wang}, \citenamefont {Bansil},
  \citenamefont {Chou}, \citenamefont {Shibayev}, \citenamefont {Lin},
  \citenamefont {Jia},\ and\ \citenamefont {Hasan}}]{xu2015discovery}%
  \BibitemOpen
  \bibfield  {author} {\bibinfo {author} {\bibfnamefont {S.-Y.}\ \bibnamefont
  {Xu}}, \bibinfo {author} {\bibfnamefont {I.}~\bibnamefont {Belopolski}},
  \bibinfo {author} {\bibfnamefont {N.}~\bibnamefont {Alidoust}}, \bibinfo
  {author} {\bibfnamefont {M.}~\bibnamefont {Neupane}}, \bibinfo {author}
  {\bibfnamefont {G.}~\bibnamefont {Bian}}, \bibinfo {author} {\bibfnamefont
  {C.}~\bibnamefont {Zhang}}, \bibinfo {author} {\bibfnamefont
  {R.}~\bibnamefont {Sankar}}, \bibinfo {author} {\bibfnamefont
  {G.}~\bibnamefont {Chang}}, \bibinfo {author} {\bibfnamefont
  {Z.}~\bibnamefont {Yuan}}, \bibinfo {author} {\bibfnamefont {C.-C.}\
  \bibnamefont {Lee}}, \bibinfo {author} {\bibfnamefont {S.-M.}\ \bibnamefont
  {Huang}}, \bibinfo {author} {\bibfnamefont {H.}~\bibnamefont {Zheng}},
  \bibinfo {author} {\bibfnamefont {J.}~\bibnamefont {Ma}}, \bibinfo {author}
  {\bibfnamefont {D.~S.}\ \bibnamefont {Sanchez}}, \bibinfo {author}
  {\bibfnamefont {B.}~\bibnamefont {Wang}}, \bibinfo {author} {\bibfnamefont
  {A.}~\bibnamefont {Bansil}}, \bibinfo {author} {\bibfnamefont
  {F.}~\bibnamefont {Chou}}, \bibinfo {author} {\bibfnamefont {P.~P.}\
  \bibnamefont {Shibayev}}, \bibinfo {author} {\bibfnamefont {H.}~\bibnamefont
  {Lin}}, \bibinfo {author} {\bibfnamefont {S.}~\bibnamefont {Jia}},\ and\
  \bibinfo {author} {\bibfnamefont {M.~Z.}\ \bibnamefont {Hasan}},\ }\bibfield
  {title} {\bibinfo {title} {{Discovery of a Weyl fermion semimetal and
  topological Fermi arcs}},\ }\href {https://doi.org/10.1126/science.aaa9297}
  {\bibfield  {journal} {\bibinfo  {journal} {Science}\ }\textbf {\bibinfo
  {volume} {349}},\ \bibinfo {pages} {613} (\bibinfo {year}
  {2015}{\natexlab{a}})}\BibitemShut {NoStop}%
\bibitem [{\citenamefont {Zhang}\ \emph {et~al.}(2016)\citenamefont {Zhang},
  \citenamefont {Xu}, \citenamefont {Belopolski}, \citenamefont {Yuan},
  \citenamefont {Lin}, \citenamefont {Tong}, \citenamefont {Bian},
  \citenamefont {Alidoust}, \citenamefont {Lee}, \citenamefont {Huang} \emph
  {et~al.}}]{zhang2016signatures}%
  \BibitemOpen
  \bibfield  {author} {\bibinfo {author} {\bibfnamefont {C.-L.}\ \bibnamefont
  {Zhang}}, \bibinfo {author} {\bibfnamefont {S.-Y.}\ \bibnamefont {Xu}},
  \bibinfo {author} {\bibfnamefont {I.}~\bibnamefont {Belopolski}}, \bibinfo
  {author} {\bibfnamefont {Z.}~\bibnamefont {Yuan}}, \bibinfo {author}
  {\bibfnamefont {Z.}~\bibnamefont {Lin}}, \bibinfo {author} {\bibfnamefont
  {B.}~\bibnamefont {Tong}}, \bibinfo {author} {\bibfnamefont {G.}~\bibnamefont
  {Bian}}, \bibinfo {author} {\bibfnamefont {N.}~\bibnamefont {Alidoust}},
  \bibinfo {author} {\bibfnamefont {C.-C.}\ \bibnamefont {Lee}}, \bibinfo
  {author} {\bibfnamefont {S.-M.}\ \bibnamefont {Huang}}, \emph {et~al.},\
  }\bibfield  {title} {\bibinfo {title} {{Signatures of the Adler--Bell--Jackiw
  chiral anomaly in a Weyl fermion semimetal}},\ }\href
  {https://doi.org/10.1038/ncomms10735} {\bibfield  {journal} {\bibinfo
  {journal} {Nat. Commun.}\ }\textbf {\bibinfo {volume} {7}},\ \bibinfo {pages}
  {1} (\bibinfo {year} {2016})}\BibitemShut {NoStop}%
\bibitem [{\citenamefont {Huang}\ \emph
  {et~al.}(2015{\natexlab{b}})\citenamefont {Huang}, \citenamefont {Zhao},
  \citenamefont {Long}, \citenamefont {Wang}, \citenamefont {Chen},
  \citenamefont {Yang}, \citenamefont {Liang}, \citenamefont {Xue},
  \citenamefont {Weng}, \citenamefont {Fang}, \citenamefont {Dai},\ and\
  \citenamefont {Chen}}]{huang2015observation}%
  \BibitemOpen
  \bibfield  {author} {\bibinfo {author} {\bibfnamefont {X.}~\bibnamefont
  {Huang}}, \bibinfo {author} {\bibfnamefont {L.}~\bibnamefont {Zhao}},
  \bibinfo {author} {\bibfnamefont {Y.}~\bibnamefont {Long}}, \bibinfo {author}
  {\bibfnamefont {P.}~\bibnamefont {Wang}}, \bibinfo {author} {\bibfnamefont
  {D.}~\bibnamefont {Chen}}, \bibinfo {author} {\bibfnamefont {Z.}~\bibnamefont
  {Yang}}, \bibinfo {author} {\bibfnamefont {H.}~\bibnamefont {Liang}},
  \bibinfo {author} {\bibfnamefont {M.}~\bibnamefont {Xue}}, \bibinfo {author}
  {\bibfnamefont {H.}~\bibnamefont {Weng}}, \bibinfo {author} {\bibfnamefont
  {Z.}~\bibnamefont {Fang}}, \bibinfo {author} {\bibfnamefont {X.}~\bibnamefont
  {Dai}},\ and\ \bibinfo {author} {\bibfnamefont {G.}~\bibnamefont {Chen}},\
  }\bibfield  {title} {\bibinfo {title} {{Observation of the
  Chiral-Anomaly-Induced Negative Magnetoresistance in 3D Weyl Semimetal
  TaAs}},\ }\href {https://doi.org/10.1103/PhysRevX.5.031023} {\bibfield
  {journal} {\bibinfo  {journal} {Phys. Rev. X}\ }\textbf {\bibinfo {volume}
  {5}},\ \bibinfo {pages} {031023} (\bibinfo {year}
  {2015}{\natexlab{b}})}\BibitemShut {NoStop}%
\bibitem [{\citenamefont {Shekhar}\ \emph {et~al.}(2015)\citenamefont
  {Shekhar}, \citenamefont {Nayak}, \citenamefont {Sun}, \citenamefont
  {Schmidt}, \citenamefont {Nicklas}, \citenamefont {Leermakers}, \citenamefont
  {Zeitler}, \citenamefont {Skourski}, \citenamefont {Wosnitza}, \citenamefont
  {Liu} \emph {et~al.}}]{shekhar2015extremely}%
  \BibitemOpen
  \bibfield  {author} {\bibinfo {author} {\bibfnamefont {C.}~\bibnamefont
  {Shekhar}}, \bibinfo {author} {\bibfnamefont {A.~K.}\ \bibnamefont {Nayak}},
  \bibinfo {author} {\bibfnamefont {Y.}~\bibnamefont {Sun}}, \bibinfo {author}
  {\bibfnamefont {M.}~\bibnamefont {Schmidt}}, \bibinfo {author} {\bibfnamefont
  {M.}~\bibnamefont {Nicklas}}, \bibinfo {author} {\bibfnamefont
  {I.}~\bibnamefont {Leermakers}}, \bibinfo {author} {\bibfnamefont
  {U.}~\bibnamefont {Zeitler}}, \bibinfo {author} {\bibfnamefont
  {Y.}~\bibnamefont {Skourski}}, \bibinfo {author} {\bibfnamefont
  {J.}~\bibnamefont {Wosnitza}}, \bibinfo {author} {\bibfnamefont
  {Z.}~\bibnamefont {Liu}}, \emph {et~al.},\ }\bibfield  {title} {\bibinfo
  {title} {{Extremely large magnetoresistance and ultrahigh mobility in the
  topological Weyl semimetal candidate NbP}},\ }\href
  {https://doi.org/10.1038/nphys3372} {\bibfield  {journal} {\bibinfo
  {journal} {Nat. Phys.}\ }\textbf {\bibinfo {volume} {11}},\ \bibinfo {pages}
  {645} (\bibinfo {year} {2015})}\BibitemShut {NoStop}%
\bibitem [{\citenamefont {Ali}\ \emph {et~al.}(2014)\citenamefont {Ali},
  \citenamefont {Xiong}, \citenamefont {Flynn}, \citenamefont {Tao},
  \citenamefont {Gibson}, \citenamefont {Schoop}, \citenamefont {Liang},
  \citenamefont {Haldolaarachchige}, \citenamefont {Hirschberger},
  \citenamefont {Ong} \emph {et~al.}}]{ali2014large}%
  \BibitemOpen
  \bibfield  {author} {\bibinfo {author} {\bibfnamefont {M.~N.}\ \bibnamefont
  {Ali}}, \bibinfo {author} {\bibfnamefont {J.}~\bibnamefont {Xiong}}, \bibinfo
  {author} {\bibfnamefont {S.}~\bibnamefont {Flynn}}, \bibinfo {author}
  {\bibfnamefont {J.}~\bibnamefont {Tao}}, \bibinfo {author} {\bibfnamefont
  {Q.~D.}\ \bibnamefont {Gibson}}, \bibinfo {author} {\bibfnamefont {L.~M.}\
  \bibnamefont {Schoop}}, \bibinfo {author} {\bibfnamefont {T.}~\bibnamefont
  {Liang}}, \bibinfo {author} {\bibfnamefont {N.}~\bibnamefont
  {Haldolaarachchige}}, \bibinfo {author} {\bibfnamefont {M.}~\bibnamefont
  {Hirschberger}}, \bibinfo {author} {\bibfnamefont {N.~P.}\ \bibnamefont
  {Ong}}, \emph {et~al.},\ }\bibfield  {title} {\bibinfo {title} {{Large,
  non-saturating magnetoresistance in WTe$_{2}$}},\ }\href
  {https://doi.org/10.1038/nature13763} {\bibfield  {journal} {\bibinfo
  {journal} {Nature}\ }\textbf {\bibinfo {volume} {514}},\ \bibinfo {pages}
  {205} (\bibinfo {year} {2014})}\BibitemShut {NoStop}%
\bibitem [{\citenamefont {Zhu}\ \emph {et~al.}(2015)\citenamefont {Zhu},
  \citenamefont {Lin}, \citenamefont {Liu}, \citenamefont {Fauqu\'e},
  \citenamefont {Tao}, \citenamefont {Yang}, \citenamefont {Shi},\ and\
  \citenamefont {Behnia}}]{zhu2015quantum}%
  \BibitemOpen
  \bibfield  {author} {\bibinfo {author} {\bibfnamefont {Z.}~\bibnamefont
  {Zhu}}, \bibinfo {author} {\bibfnamefont {X.}~\bibnamefont {Lin}}, \bibinfo
  {author} {\bibfnamefont {J.}~\bibnamefont {Liu}}, \bibinfo {author}
  {\bibfnamefont {B.}~\bibnamefont {Fauqu\'e}}, \bibinfo {author}
  {\bibfnamefont {Q.}~\bibnamefont {Tao}}, \bibinfo {author} {\bibfnamefont
  {C.}~\bibnamefont {Yang}}, \bibinfo {author} {\bibfnamefont {Y.}~\bibnamefont
  {Shi}},\ and\ \bibinfo {author} {\bibfnamefont {K.}~\bibnamefont {Behnia}},\
  }\bibfield  {title} {\bibinfo {title} {{Quantum Oscillations, Thermoelectric
  Coefficients, and the Fermi Surface of Semimetallic ${\mathrm{WTe}}_{2}$}},\
  }\href {https://doi.org/10.1103/PhysRevLett.114.176601} {\bibfield  {journal}
  {\bibinfo  {journal} {Phys. Rev. Lett.}\ }\textbf {\bibinfo {volume} {114}},\
  \bibinfo {pages} {176601} (\bibinfo {year} {2015})}\BibitemShut {NoStop}%
\bibitem [{\citenamefont {Jiang}\ \emph {et~al.}(2017)\citenamefont {Jiang},
  \citenamefont {Liu}, \citenamefont {Sun}, \citenamefont {Yang}, \citenamefont
  {Rajamathi}, \citenamefont {Qi}, \citenamefont {Yang}, \citenamefont {Chen},
  \citenamefont {Peng}, \citenamefont {Hwang} \emph
  {et~al.}}]{jiang2017signature}%
  \BibitemOpen
  \bibfield  {author} {\bibinfo {author} {\bibfnamefont {J.}~\bibnamefont
  {Jiang}}, \bibinfo {author} {\bibfnamefont {Z.}~\bibnamefont {Liu}}, \bibinfo
  {author} {\bibfnamefont {Y.}~\bibnamefont {Sun}}, \bibinfo {author}
  {\bibfnamefont {H.}~\bibnamefont {Yang}}, \bibinfo {author} {\bibfnamefont
  {C.}~\bibnamefont {Rajamathi}}, \bibinfo {author} {\bibfnamefont
  {Y.}~\bibnamefont {Qi}}, \bibinfo {author} {\bibfnamefont {L.}~\bibnamefont
  {Yang}}, \bibinfo {author} {\bibfnamefont {C.}~\bibnamefont {Chen}}, \bibinfo
  {author} {\bibfnamefont {H.}~\bibnamefont {Peng}}, \bibinfo {author}
  {\bibfnamefont {C.}~\bibnamefont {Hwang}}, \emph {et~al.},\ }\bibfield
  {title} {\bibinfo {title} {{Signature of type-II Weyl semimetal phase in
  MoTe$_{2}$}},\ }\href {https://doi.org/10.1038/ncomms13973} {\bibfield
  {journal} {\bibinfo  {journal} {Nat. Commun.}\ }\textbf {\bibinfo {volume}
  {8}},\ \bibinfo {pages} {1} (\bibinfo {year} {2017})}\BibitemShut {NoStop}%
\bibitem [{\citenamefont {Kuroda}\ \emph {et~al.}(2017)\citenamefont {Kuroda},
  \citenamefont {Tomita}, \citenamefont {Suzuki}, \citenamefont {Bareille},
  \citenamefont {Nugroho}, \citenamefont {Goswami}, \citenamefont {Ochi},
  \citenamefont {Ikhlas}, \citenamefont {Nakayama}, \citenamefont {Akebi} \emph
  {et~al.}}]{kuroda2017evidence}%
  \BibitemOpen
  \bibfield  {author} {\bibinfo {author} {\bibfnamefont {K.}~\bibnamefont
  {Kuroda}}, \bibinfo {author} {\bibfnamefont {T.}~\bibnamefont {Tomita}},
  \bibinfo {author} {\bibfnamefont {M.-T.}\ \bibnamefont {Suzuki}}, \bibinfo
  {author} {\bibfnamefont {C.}~\bibnamefont {Bareille}}, \bibinfo {author}
  {\bibfnamefont {A.}~\bibnamefont {Nugroho}}, \bibinfo {author} {\bibfnamefont
  {P.}~\bibnamefont {Goswami}}, \bibinfo {author} {\bibfnamefont
  {M.}~\bibnamefont {Ochi}}, \bibinfo {author} {\bibfnamefont {M.}~\bibnamefont
  {Ikhlas}}, \bibinfo {author} {\bibfnamefont {M.}~\bibnamefont {Nakayama}},
  \bibinfo {author} {\bibfnamefont {S.}~\bibnamefont {Akebi}}, \emph {et~al.},\
  }\bibfield  {title} {\bibinfo {title} {{Evidence for magnetic Weyl fermions
  in a correlated metal}},\ }\href {https://doi.org/10.1038/nmat4987}
  {\bibfield  {journal} {\bibinfo  {journal} {Nat. Mater.}\ }\textbf {\bibinfo
  {volume} {16}},\ \bibinfo {pages} {1090} (\bibinfo {year}
  {2017})}\BibitemShut {NoStop}%
\bibitem [{\citenamefont {Yang}\ \emph {et~al.}(2017)\citenamefont {Yang},
  \citenamefont {Sun}, \citenamefont {Zhang}, \citenamefont {Shi},
  \citenamefont {Parkin},\ and\ \citenamefont {Yan}}]{yang2017topological}%
  \BibitemOpen
  \bibfield  {author} {\bibinfo {author} {\bibfnamefont {H.}~\bibnamefont
  {Yang}}, \bibinfo {author} {\bibfnamefont {Y.}~\bibnamefont {Sun}}, \bibinfo
  {author} {\bibfnamefont {Y.}~\bibnamefont {Zhang}}, \bibinfo {author}
  {\bibfnamefont {W.-J.}\ \bibnamefont {Shi}}, \bibinfo {author} {\bibfnamefont
  {S.~S.~P.}\ \bibnamefont {Parkin}},\ and\ \bibinfo {author} {\bibfnamefont
  {B.}~\bibnamefont {Yan}},\ }\bibfield  {title} {\bibinfo {title}
  {{Topological Weyl semimetals in the chiral antiferromagnetic materials
  Mn$_{3}$Ge and Mn$_{3}$Sn}},\ }\href
  {https://doi.org/10.1088/1367-2630/aa5487} {\bibfield  {journal} {\bibinfo
  {journal} {New J. Phys.}\ }\textbf {\bibinfo {volume} {19}},\ \bibinfo
  {pages} {015008} (\bibinfo {year} {2017})}\BibitemShut {NoStop}%
\bibitem [{\citenamefont {Morali}\ \emph {et~al.}(2019)\citenamefont {Morali},
  \citenamefont {Batabyal}, \citenamefont {Nag}, \citenamefont {Liu},
  \citenamefont {Xu}, \citenamefont {Sun}, \citenamefont {Yan}, \citenamefont
  {Felser}, \citenamefont {Avraham},\ and\ \citenamefont
  {Beidenkopf}}]{morali2019fermi}%
  \BibitemOpen
  \bibfield  {author} {\bibinfo {author} {\bibfnamefont {N.}~\bibnamefont
  {Morali}}, \bibinfo {author} {\bibfnamefont {R.}~\bibnamefont {Batabyal}},
  \bibinfo {author} {\bibfnamefont {P.~K.}\ \bibnamefont {Nag}}, \bibinfo
  {author} {\bibfnamefont {E.}~\bibnamefont {Liu}}, \bibinfo {author}
  {\bibfnamefont {Q.}~\bibnamefont {Xu}}, \bibinfo {author} {\bibfnamefont
  {Y.}~\bibnamefont {Sun}}, \bibinfo {author} {\bibfnamefont {B.}~\bibnamefont
  {Yan}}, \bibinfo {author} {\bibfnamefont {C.}~\bibnamefont {Felser}},
  \bibinfo {author} {\bibfnamefont {N.}~\bibnamefont {Avraham}},\ and\ \bibinfo
  {author} {\bibfnamefont {H.}~\bibnamefont {Beidenkopf}},\ }\bibfield  {title}
  {\bibinfo {title} {{Fermi-arc diversity on surface terminations of the
  magnetic Weyl semimetal Co$_{3}$Sn$_{2}$S$_{2}$}},\ }\href
  {https://doi.org/10.1126/science.aav2334} {\bibfield  {journal} {\bibinfo
  {journal} {Science}\ }\textbf {\bibinfo {volume} {365}},\ \bibinfo {pages}
  {1286} (\bibinfo {year} {2019})}\BibitemShut {NoStop}%
\bibitem [{\citenamefont {Liu}\ \emph {et~al.}(2019)\citenamefont {Liu},
  \citenamefont {Liang}, \citenamefont {Liu}, \citenamefont {Xu}, \citenamefont
  {Li}, \citenamefont {Chen}, \citenamefont {Pei}, \citenamefont {Shi},
  \citenamefont {Mo}, \citenamefont {Dudin}, \citenamefont {Kim}, \citenamefont
  {Cacho}, \citenamefont {Li}, \citenamefont {Sun}, \citenamefont {Yang},
  \citenamefont {Liu}, \citenamefont {Parkin}, \citenamefont {Felser},\ and\
  \citenamefont {Chen}}]{liu2019magnetic}%
  \BibitemOpen
  \bibfield  {author} {\bibinfo {author} {\bibfnamefont {D.~F.}\ \bibnamefont
  {Liu}}, \bibinfo {author} {\bibfnamefont {A.~J.}\ \bibnamefont {Liang}},
  \bibinfo {author} {\bibfnamefont {E.~K.}\ \bibnamefont {Liu}}, \bibinfo
  {author} {\bibfnamefont {Q.~N.}\ \bibnamefont {Xu}}, \bibinfo {author}
  {\bibfnamefont {Y.~W.}\ \bibnamefont {Li}}, \bibinfo {author} {\bibfnamefont
  {C.}~\bibnamefont {Chen}}, \bibinfo {author} {\bibfnamefont {D.}~\bibnamefont
  {Pei}}, \bibinfo {author} {\bibfnamefont {W.~J.}\ \bibnamefont {Shi}},
  \bibinfo {author} {\bibfnamefont {S.~K.}\ \bibnamefont {Mo}}, \bibinfo
  {author} {\bibfnamefont {P.}~\bibnamefont {Dudin}}, \bibinfo {author}
  {\bibfnamefont {T.}~\bibnamefont {Kim}}, \bibinfo {author} {\bibfnamefont
  {C.}~\bibnamefont {Cacho}}, \bibinfo {author} {\bibfnamefont
  {G.}~\bibnamefont {Li}}, \bibinfo {author} {\bibfnamefont {Y.}~\bibnamefont
  {Sun}}, \bibinfo {author} {\bibfnamefont {L.~X.}\ \bibnamefont {Yang}},
  \bibinfo {author} {\bibfnamefont {Z.~K.}\ \bibnamefont {Liu}}, \bibinfo
  {author} {\bibfnamefont {S.~S.~P.}\ \bibnamefont {Parkin}}, \bibinfo {author}
  {\bibfnamefont {C.}~\bibnamefont {Felser}},\ and\ \bibinfo {author}
  {\bibfnamefont {Y.~L.}\ \bibnamefont {Chen}},\ }\bibfield  {title} {\bibinfo
  {title} {{Magnetic Weyl semimetal phase in a Kagom\'e crystal}},\ }\href
  {https://doi.org/10.1126/science.aav2873} {\bibfield  {journal} {\bibinfo
  {journal} {Science}\ }\textbf {\bibinfo {volume} {365}},\ \bibinfo {pages}
  {1282} (\bibinfo {year} {2019})}\BibitemShut {NoStop}%
\bibitem [{\citenamefont {Belopolski}\ \emph {et~al.}(2019)\citenamefont
  {Belopolski}, \citenamefont {Manna}, \citenamefont {Sanchez}, \citenamefont
  {Chang}, \citenamefont {Ernst}, \citenamefont {Yin}, \citenamefont {Zhang},
  \citenamefont {Cochran}, \citenamefont {Shumiya}, \citenamefont {Zheng},
  \citenamefont {Singh}, \citenamefont {Bian}, \citenamefont {Multer},
  \citenamefont {Litskevich}, \citenamefont {Zhou}, \citenamefont {Huang},
  \citenamefont {Wang}, \citenamefont {Chang}, \citenamefont {Xu},
  \citenamefont {Bansil}, \citenamefont {Felser}, \citenamefont {Lin},\ and\
  \citenamefont {Hasan}}]{belopolski2019discovery}%
  \BibitemOpen
  \bibfield  {author} {\bibinfo {author} {\bibfnamefont {I.}~\bibnamefont
  {Belopolski}}, \bibinfo {author} {\bibfnamefont {K.}~\bibnamefont {Manna}},
  \bibinfo {author} {\bibfnamefont {D.~S.}\ \bibnamefont {Sanchez}}, \bibinfo
  {author} {\bibfnamefont {G.}~\bibnamefont {Chang}}, \bibinfo {author}
  {\bibfnamefont {B.}~\bibnamefont {Ernst}}, \bibinfo {author} {\bibfnamefont
  {J.}~\bibnamefont {Yin}}, \bibinfo {author} {\bibfnamefont {S.~S.}\
  \bibnamefont {Zhang}}, \bibinfo {author} {\bibfnamefont {T.}~\bibnamefont
  {Cochran}}, \bibinfo {author} {\bibfnamefont {N.}~\bibnamefont {Shumiya}},
  \bibinfo {author} {\bibfnamefont {H.}~\bibnamefont {Zheng}}, \bibinfo
  {author} {\bibfnamefont {B.}~\bibnamefont {Singh}}, \bibinfo {author}
  {\bibfnamefont {G.}~\bibnamefont {Bian}}, \bibinfo {author} {\bibfnamefont
  {D.}~\bibnamefont {Multer}}, \bibinfo {author} {\bibfnamefont
  {M.}~\bibnamefont {Litskevich}}, \bibinfo {author} {\bibfnamefont
  {X.}~\bibnamefont {Zhou}}, \bibinfo {author} {\bibfnamefont {S.-M.}\
  \bibnamefont {Huang}}, \bibinfo {author} {\bibfnamefont {B.}~\bibnamefont
  {Wang}}, \bibinfo {author} {\bibfnamefont {T.-R.}\ \bibnamefont {Chang}},
  \bibinfo {author} {\bibfnamefont {S.-Y.}\ \bibnamefont {Xu}}, \bibinfo
  {author} {\bibfnamefont {A.}~\bibnamefont {Bansil}}, \bibinfo {author}
  {\bibfnamefont {C.}~\bibnamefont {Felser}}, \bibinfo {author} {\bibfnamefont
  {H.}~\bibnamefont {Lin}},\ and\ \bibinfo {author} {\bibfnamefont {M.~Z.}\
  \bibnamefont {Hasan}},\ }\bibfield  {title} {\bibinfo {title} {{Discovery of
  topological Weyl fermion lines and drumhead surface states in a room
  temperature magnet}},\ }\href {https://doi.org/10.1126/science.aav2327}
  {\bibfield  {journal} {\bibinfo  {journal} {Science}\ }\textbf {\bibinfo
  {volume} {365}},\ \bibinfo {pages} {1278} (\bibinfo {year}
  {2019})}\BibitemShut {NoStop}%
\bibitem [{\citenamefont {Borisenko}\ \emph {et~al.}(2019)\citenamefont
  {Borisenko}, \citenamefont {Evtushinsky}, \citenamefont {Gibson},
  \citenamefont {Yaresko}, \citenamefont {Koepernik}, \citenamefont {Kim},
  \citenamefont {Ali}, \citenamefont {van~den Brink}, \citenamefont {Hoesch},
  \citenamefont {Fedorov} \emph {et~al.}}]{borisenko2019time}%
  \BibitemOpen
  \bibfield  {author} {\bibinfo {author} {\bibfnamefont {S.}~\bibnamefont
  {Borisenko}}, \bibinfo {author} {\bibfnamefont {D.}~\bibnamefont
  {Evtushinsky}}, \bibinfo {author} {\bibfnamefont {Q.}~\bibnamefont {Gibson}},
  \bibinfo {author} {\bibfnamefont {A.}~\bibnamefont {Yaresko}}, \bibinfo
  {author} {\bibfnamefont {K.}~\bibnamefont {Koepernik}}, \bibinfo {author}
  {\bibfnamefont {T.}~\bibnamefont {Kim}}, \bibinfo {author} {\bibfnamefont
  {M.}~\bibnamefont {Ali}}, \bibinfo {author} {\bibfnamefont {J.}~\bibnamefont
  {van~den Brink}}, \bibinfo {author} {\bibfnamefont {M.}~\bibnamefont
  {Hoesch}}, \bibinfo {author} {\bibfnamefont {A.}~\bibnamefont {Fedorov}},
  \emph {et~al.},\ }\bibfield  {title} {\bibinfo {title} {{Time-reversal
  symmetry breaking type-II Weyl state in YbMnBi$_{2}$}},\ }\href
  {https://doi.org/10.1038/s41467-019-11393-5} {\bibfield  {journal} {\bibinfo
  {journal} {Nat. Commun.}\ }\textbf {\bibinfo {volume} {10}},\ \bibinfo
  {pages} {1} (\bibinfo {year} {2019})}\BibitemShut {NoStop}%
\bibitem [{\citenamefont {Kurebayashi}\ and\ \citenamefont
  {Nomura}(2016)}]{kurebayashi2016voltage}%
  \BibitemOpen
  \bibfield  {author} {\bibinfo {author} {\bibfnamefont {D.}~\bibnamefont
  {Kurebayashi}}\ and\ \bibinfo {author} {\bibfnamefont {K.}~\bibnamefont
  {Nomura}},\ }\bibfield  {title} {\bibinfo {title} {{Voltage-Driven
  Magnetization Switching and Spin Pumping in Weyl Semimetals}},\ }\href
  {https://doi.org/10.1103/PhysRevApplied.6.044013} {\bibfield  {journal}
  {\bibinfo  {journal} {Phys. Rev. Appl.}\ }\textbf {\bibinfo {volume} {6}},\
  \bibinfo {pages} {044013} (\bibinfo {year} {2016})}\BibitemShut {NoStop}%
\bibitem [{\citenamefont {Yang}\ \emph
  {et~al.}(2021{\natexlab{a}})\citenamefont {Yang}, \citenamefont {Naaman},
  \citenamefont {Paltiel},\ and\ \citenamefont {Parkin}}]{yang2021chiral}%
  \BibitemOpen
  \bibfield  {author} {\bibinfo {author} {\bibfnamefont {S.-H.}\ \bibnamefont
  {Yang}}, \bibinfo {author} {\bibfnamefont {R.}~\bibnamefont {Naaman}},
  \bibinfo {author} {\bibfnamefont {Y.}~\bibnamefont {Paltiel}},\ and\ \bibinfo
  {author} {\bibfnamefont {S.~S.}\ \bibnamefont {Parkin}},\ }\bibfield  {title}
  {\bibinfo {title} {Chiral spintronics},\ }\href
  {https://doi.org/10.1038/s42254-021-00302-9} {\bibfield  {journal} {\bibinfo
  {journal} {Nature Reviews Physics}\ }\textbf {\bibinfo {volume} {3}},\
  \bibinfo {pages} {328} (\bibinfo {year} {2021}{\natexlab{a}})}\BibitemShut
  {NoStop}%
\bibitem [{\citenamefont {Grefe}\ \emph {et~al.}(2020)\citenamefont {Grefe},
  \citenamefont {Lai}, \citenamefont {Paschen},\ and\ \citenamefont
  {Si}}]{grefe2020weyl}%
  \BibitemOpen
  \bibfield  {author} {\bibinfo {author} {\bibfnamefont {S.~E.}\ \bibnamefont
  {Grefe}}, \bibinfo {author} {\bibfnamefont {H.-H.}\ \bibnamefont {Lai}},
  \bibinfo {author} {\bibfnamefont {S.}~\bibnamefont {Paschen}},\ and\ \bibinfo
  {author} {\bibfnamefont {Q.}~\bibnamefont {Si}},\ }\bibfield  {title}
  {\bibinfo {title} {{Weyl-Kondo semimetals in nonsymmorphic systems}},\ }\href
  {https://doi.org/10.1103/PhysRevB.101.075138} {\bibfield  {journal} {\bibinfo
   {journal} {Phys. Rev. B}\ }\textbf {\bibinfo {volume} {101}},\ \bibinfo
  {pages} {075138} (\bibinfo {year} {2020})}\BibitemShut {NoStop}%
\bibitem [{\citenamefont {Paschen}\ and\ \citenamefont
  {Si}(2021)}]{paschen2021quantum}%
  \BibitemOpen
  \bibfield  {author} {\bibinfo {author} {\bibfnamefont {S.}~\bibnamefont
  {Paschen}}\ and\ \bibinfo {author} {\bibfnamefont {Q.}~\bibnamefont {Si}},\
  }\bibfield  {title} {\bibinfo {title} {Quantum phases driven by strong
  correlations},\ }\href {https://doi.org/10.1038/s42254-020-00262-6}
  {\bibfield  {journal} {\bibinfo  {journal} {Nat. Rev. Phys.}\ }\textbf
  {\bibinfo {volume} {3}},\ \bibinfo {pages} {9} (\bibinfo {year}
  {2021})}\BibitemShut {NoStop}%
\bibitem [{\citenamefont {Xu}\ \emph {et~al.}(2015{\natexlab{b}})\citenamefont
  {Xu}, \citenamefont {Belopolski}, \citenamefont {Sanchez}, \citenamefont
  {Zhang}, \citenamefont {Chang}, \citenamefont {Guo}, \citenamefont {Bian},
  \citenamefont {Yuan}, \citenamefont {Lu}, \citenamefont {Chang},
  \citenamefont {Shibayev}, \citenamefont {Prokopovych}, \citenamefont
  {Alidoust}, \citenamefont {Zheng}, \citenamefont {Lee}, \citenamefont
  {Huang}, \citenamefont {Sankar}, \citenamefont {Chou}, \citenamefont {Hsu},
  \citenamefont {Jeng}, \citenamefont {Bansil}, \citenamefont {Neupert},
  \citenamefont {Strocov}, \citenamefont {Lin}, \citenamefont {Jia},\ and\
  \citenamefont {Hasan}}]{xu2015experimental}%
  \BibitemOpen
  \bibfield  {author} {\bibinfo {author} {\bibfnamefont {S.-Y.}\ \bibnamefont
  {Xu}}, \bibinfo {author} {\bibfnamefont {I.}~\bibnamefont {Belopolski}},
  \bibinfo {author} {\bibfnamefont {D.~S.}\ \bibnamefont {Sanchez}}, \bibinfo
  {author} {\bibfnamefont {C.}~\bibnamefont {Zhang}}, \bibinfo {author}
  {\bibfnamefont {G.}~\bibnamefont {Chang}}, \bibinfo {author} {\bibfnamefont
  {C.}~\bibnamefont {Guo}}, \bibinfo {author} {\bibfnamefont {G.}~\bibnamefont
  {Bian}}, \bibinfo {author} {\bibfnamefont {Z.}~\bibnamefont {Yuan}}, \bibinfo
  {author} {\bibfnamefont {H.}~\bibnamefont {Lu}}, \bibinfo {author}
  {\bibfnamefont {T.-R.}\ \bibnamefont {Chang}}, \bibinfo {author}
  {\bibfnamefont {P.~P.}\ \bibnamefont {Shibayev}}, \bibinfo {author}
  {\bibfnamefont {M.~L.}\ \bibnamefont {Prokopovych}}, \bibinfo {author}
  {\bibfnamefont {N.}~\bibnamefont {Alidoust}}, \bibinfo {author}
  {\bibfnamefont {H.}~\bibnamefont {Zheng}}, \bibinfo {author} {\bibfnamefont
  {C.-C.}\ \bibnamefont {Lee}}, \bibinfo {author} {\bibfnamefont {S.-M.}\
  \bibnamefont {Huang}}, \bibinfo {author} {\bibfnamefont {R.}~\bibnamefont
  {Sankar}}, \bibinfo {author} {\bibfnamefont {F.}~\bibnamefont {Chou}},
  \bibinfo {author} {\bibfnamefont {C.-H.}\ \bibnamefont {Hsu}}, \bibinfo
  {author} {\bibfnamefont {H.-T.}\ \bibnamefont {Jeng}}, \bibinfo {author}
  {\bibfnamefont {A.}~\bibnamefont {Bansil}}, \bibinfo {author} {\bibfnamefont
  {T.}~\bibnamefont {Neupert}}, \bibinfo {author} {\bibfnamefont {V.~N.}\
  \bibnamefont {Strocov}}, \bibinfo {author} {\bibfnamefont {H.}~\bibnamefont
  {Lin}}, \bibinfo {author} {\bibfnamefont {S.}~\bibnamefont {Jia}},\ and\
  \bibinfo {author} {\bibfnamefont {M.~Z.}\ \bibnamefont {Hasan}},\ }\bibfield
  {title} {\bibinfo {title} {{Experimental discovery of a topological Weyl
  semimetal state in TaP}},\ }\href {https://doi.org/10.1126/sciadv.1501092}
  {\bibfield  {journal} {\bibinfo  {journal} {Sci. Adv.}\ }\textbf {\bibinfo
  {volume} {1}},\ \bibinfo {pages} {e1501092} (\bibinfo {year}
  {2015}{\natexlab{b}})}\BibitemShut {NoStop}%
\bibitem [{\citenamefont {Yang}\ \emph {et~al.}(2015)\citenamefont {Yang},
  \citenamefont {Liu}, \citenamefont {Sun}, \citenamefont {Peng}, \citenamefont
  {Yang}, \citenamefont {Zhang}, \citenamefont {Zhou}, \citenamefont {Zhang},
  \citenamefont {Guo}, \citenamefont {Rahn} \emph {et~al.}}]{yang2015weyl}%
  \BibitemOpen
  \bibfield  {author} {\bibinfo {author} {\bibfnamefont {L.}~\bibnamefont
  {Yang}}, \bibinfo {author} {\bibfnamefont {Z.}~\bibnamefont {Liu}}, \bibinfo
  {author} {\bibfnamefont {Y.}~\bibnamefont {Sun}}, \bibinfo {author}
  {\bibfnamefont {H.}~\bibnamefont {Peng}}, \bibinfo {author} {\bibfnamefont
  {H.}~\bibnamefont {Yang}}, \bibinfo {author} {\bibfnamefont {T.}~\bibnamefont
  {Zhang}}, \bibinfo {author} {\bibfnamefont {B.}~\bibnamefont {Zhou}},
  \bibinfo {author} {\bibfnamefont {Y.}~\bibnamefont {Zhang}}, \bibinfo
  {author} {\bibfnamefont {Y.}~\bibnamefont {Guo}}, \bibinfo {author}
  {\bibfnamefont {M.}~\bibnamefont {Rahn}}, \emph {et~al.},\ }\bibfield
  {title} {\bibinfo {title} {{Weyl semimetal phase in the non-centrosymmetric
  compound TaAs}},\ }\href {https://doi.org/10.1038/nphys3425} {\bibfield
  {journal} {\bibinfo  {journal} {Nat. Phys.}\ }\textbf {\bibinfo {volume}
  {11}},\ \bibinfo {pages} {728} (\bibinfo {year} {2015})}\BibitemShut
  {NoStop}%
\bibitem [{\citenamefont {Arnold}\ \emph {et~al.}(2016)\citenamefont {Arnold},
  \citenamefont {Shekhar}, \citenamefont {Wu}, \citenamefont {Sun},
  \citenamefont {Dos~Reis}, \citenamefont {Kumar}, \citenamefont {Naumann},
  \citenamefont {Ajeesh}, \citenamefont {Schmidt}, \citenamefont {Grushin}
  \emph {et~al.}}]{arnold2016negative}%
  \BibitemOpen
  \bibfield  {author} {\bibinfo {author} {\bibfnamefont {F.}~\bibnamefont
  {Arnold}}, \bibinfo {author} {\bibfnamefont {C.}~\bibnamefont {Shekhar}},
  \bibinfo {author} {\bibfnamefont {S.-C.}\ \bibnamefont {Wu}}, \bibinfo
  {author} {\bibfnamefont {Y.}~\bibnamefont {Sun}}, \bibinfo {author}
  {\bibfnamefont {R.~D.}\ \bibnamefont {Dos~Reis}}, \bibinfo {author}
  {\bibfnamefont {N.}~\bibnamefont {Kumar}}, \bibinfo {author} {\bibfnamefont
  {M.}~\bibnamefont {Naumann}}, \bibinfo {author} {\bibfnamefont {M.~O.}\
  \bibnamefont {Ajeesh}}, \bibinfo {author} {\bibfnamefont {M.}~\bibnamefont
  {Schmidt}}, \bibinfo {author} {\bibfnamefont {A.~G.}\ \bibnamefont
  {Grushin}}, \emph {et~al.},\ }\bibfield  {title} {\bibinfo {title} {{Negative
  magnetoresistance without well-defined chirality in the Weyl semimetal
  TaP}},\ }\href {https://doi.org/10.1038/ncomms11615} {\bibfield  {journal}
  {\bibinfo  {journal} {Nat. Commun.}\ }\textbf {\bibinfo {volume} {7}},\
  \bibinfo {pages} {1} (\bibinfo {year} {2016})}\BibitemShut {NoStop}%
\bibitem [{\citenamefont {Liu}\ \emph {et~al.}(2016)\citenamefont {Liu},
  \citenamefont {Yang}, \citenamefont {Sun}, \citenamefont {Zhang},
  \citenamefont {Peng}, \citenamefont {Yang}, \citenamefont {Chen},
  \citenamefont {Zhang}, \citenamefont {Guo}, \citenamefont {Prabhakaran} \emph
  {et~al.}}]{liu2016evolution}%
  \BibitemOpen
  \bibfield  {author} {\bibinfo {author} {\bibfnamefont {Z.}~\bibnamefont
  {Liu}}, \bibinfo {author} {\bibfnamefont {L.}~\bibnamefont {Yang}}, \bibinfo
  {author} {\bibfnamefont {Y.}~\bibnamefont {Sun}}, \bibinfo {author}
  {\bibfnamefont {T.}~\bibnamefont {Zhang}}, \bibinfo {author} {\bibfnamefont
  {H.}~\bibnamefont {Peng}}, \bibinfo {author} {\bibfnamefont {H.}~\bibnamefont
  {Yang}}, \bibinfo {author} {\bibfnamefont {C.}~\bibnamefont {Chen}}, \bibinfo
  {author} {\bibfnamefont {Y.~f.}\ \bibnamefont {Zhang}}, \bibinfo {author}
  {\bibfnamefont {Y.}~\bibnamefont {Guo}}, \bibinfo {author} {\bibfnamefont
  {D.}~\bibnamefont {Prabhakaran}}, \emph {et~al.},\ }\bibfield  {title}
  {\bibinfo {title} {{Evolution of the Fermi surface of Weyl semimetals in the
  transition metal pnictide family}},\ }\href
  {https://doi.org/doi.org/10.1038/nmat4457} {\bibfield  {journal} {\bibinfo
  {journal} {Nat. Mater.}\ }\textbf {\bibinfo {volume} {15}},\ \bibinfo {pages}
  {27} (\bibinfo {year} {2016})}\BibitemShut {NoStop}%
\bibitem [{\citenamefont {Ng}\ \emph {et~al.}(2021)\citenamefont {Ng},
  \citenamefont {Luo}, \citenamefont {Yuan}, \citenamefont {Wu}, \citenamefont
  {Yang},\ and\ \citenamefont {Shen}}]{ng2021origin}%
  \BibitemOpen
  \bibfield  {author} {\bibinfo {author} {\bibfnamefont {T.}~\bibnamefont
  {Ng}}, \bibinfo {author} {\bibfnamefont {Y.}~\bibnamefont {Luo}}, \bibinfo
  {author} {\bibfnamefont {J.}~\bibnamefont {Yuan}}, \bibinfo {author}
  {\bibfnamefont {Y.}~\bibnamefont {Wu}}, \bibinfo {author} {\bibfnamefont
  {H.}~\bibnamefont {Yang}},\ and\ \bibinfo {author} {\bibfnamefont
  {L.}~\bibnamefont {Shen}},\ }\bibfield  {title} {\bibinfo {title} {{Origin
  and enhancement of the spin Hall angle in the Weyl semimetals LaAlSi and
  LaAlGe}},\ }\href {https://doi.org/10.1103/PhysRevB.104.014412} {\bibfield
  {journal} {\bibinfo  {journal} {Phys. Rev. B}\ }\textbf {\bibinfo {volume}
  {104}},\ \bibinfo {pages} {014412} (\bibinfo {year} {2021})}\BibitemShut
  {NoStop}%
\bibitem [{\citenamefont {Xu}\ \emph {et~al.}(2017)\citenamefont {Xu},
  \citenamefont {Alidoust}, \citenamefont {Chang}, \citenamefont {Lu},
  \citenamefont {Singh}, \citenamefont {Belopolski}, \citenamefont {Sanchez},
  \citenamefont {Zhang}, \citenamefont {Bian}, \citenamefont {Zheng},
  \citenamefont {Husanu}, \citenamefont {Bian}, \citenamefont {Huang},
  \citenamefont {Hsu}, \citenamefont {Chang}, \citenamefont {Jeng},
  \citenamefont {Bansil}, \citenamefont {Neupert}, \citenamefont {Strocov},
  \citenamefont {Lin}, \citenamefont {Jia},\ and\ \citenamefont
  {Hasan}}]{xu2017discovery}%
  \BibitemOpen
  \bibfield  {author} {\bibinfo {author} {\bibfnamefont {S.-Y.}\ \bibnamefont
  {Xu}}, \bibinfo {author} {\bibfnamefont {N.}~\bibnamefont {Alidoust}},
  \bibinfo {author} {\bibfnamefont {G.}~\bibnamefont {Chang}}, \bibinfo
  {author} {\bibfnamefont {H.}~\bibnamefont {Lu}}, \bibinfo {author}
  {\bibfnamefont {B.}~\bibnamefont {Singh}}, \bibinfo {author} {\bibfnamefont
  {I.}~\bibnamefont {Belopolski}}, \bibinfo {author} {\bibfnamefont {D.~S.}\
  \bibnamefont {Sanchez}}, \bibinfo {author} {\bibfnamefont {X.}~\bibnamefont
  {Zhang}}, \bibinfo {author} {\bibfnamefont {G.}~\bibnamefont {Bian}},
  \bibinfo {author} {\bibfnamefont {H.}~\bibnamefont {Zheng}}, \bibinfo
  {author} {\bibfnamefont {M.-A.}\ \bibnamefont {Husanu}}, \bibinfo {author}
  {\bibfnamefont {Y.}~\bibnamefont {Bian}}, \bibinfo {author} {\bibfnamefont
  {S.-M.}\ \bibnamefont {Huang}}, \bibinfo {author} {\bibfnamefont {C.-H.}\
  \bibnamefont {Hsu}}, \bibinfo {author} {\bibfnamefont {T.-R.}\ \bibnamefont
  {Chang}}, \bibinfo {author} {\bibfnamefont {H.-T.}\ \bibnamefont {Jeng}},
  \bibinfo {author} {\bibfnamefont {A.}~\bibnamefont {Bansil}}, \bibinfo
  {author} {\bibfnamefont {T.}~\bibnamefont {Neupert}}, \bibinfo {author}
  {\bibfnamefont {V.~N.}\ \bibnamefont {Strocov}}, \bibinfo {author}
  {\bibfnamefont {H.}~\bibnamefont {Lin}}, \bibinfo {author} {\bibfnamefont
  {S.}~\bibnamefont {Jia}},\ and\ \bibinfo {author} {\bibfnamefont {M.~Z.}\
  \bibnamefont {Hasan}},\ }\bibfield  {title} {\bibinfo {title} {{Discovery of
  Lorentz-violating type II Weyl fermions in LaAlGe}},\ }\href
  {https://doi.org/10.1126/sciadv.1603266} {\bibfield  {journal} {\bibinfo
  {journal} {Sci. Adv.}\ }\textbf {\bibinfo {volume} {3}},\ \bibinfo {pages}
  {e1603266} (\bibinfo {year} {2017})}\BibitemShut {NoStop}%
\bibitem [{\citenamefont {Su}\ \emph {et~al.}(2021)\citenamefont {Su},
  \citenamefont {Shi}, \citenamefont {Yuan}, \citenamefont {Wan}, \citenamefont
  {Cheng}, \citenamefont {Xi}, \citenamefont {Pi}, \citenamefont {Wang},
  \citenamefont {Zou}, \citenamefont {Yu}, \citenamefont {Zhao}, \citenamefont
  {Li},\ and\ \citenamefont {Guo}}]{su2021multiple}%
  \BibitemOpen
  \bibfield  {author} {\bibinfo {author} {\bibfnamefont {H.}~\bibnamefont
  {Su}}, \bibinfo {author} {\bibfnamefont {X.}~\bibnamefont {Shi}}, \bibinfo
  {author} {\bibfnamefont {J.}~\bibnamefont {Yuan}}, \bibinfo {author}
  {\bibfnamefont {Y.}~\bibnamefont {Wan}}, \bibinfo {author} {\bibfnamefont
  {E.}~\bibnamefont {Cheng}}, \bibinfo {author} {\bibfnamefont
  {C.}~\bibnamefont {Xi}}, \bibinfo {author} {\bibfnamefont {L.}~\bibnamefont
  {Pi}}, \bibinfo {author} {\bibfnamefont {X.}~\bibnamefont {Wang}}, \bibinfo
  {author} {\bibfnamefont {Z.}~\bibnamefont {Zou}}, \bibinfo {author}
  {\bibfnamefont {N.}~\bibnamefont {Yu}}, \bibinfo {author} {\bibfnamefont
  {W.}~\bibnamefont {Zhao}}, \bibinfo {author} {\bibfnamefont {S.}~\bibnamefont
  {Li}},\ and\ \bibinfo {author} {\bibfnamefont {Y.}~\bibnamefont {Guo}},\
  }\bibfield  {title} {\bibinfo {title} {{Multiple Weyl fermions in the
  noncentrosymmetric semimetal LaAlSi}},\ }\href
  {https://doi.org/10.1103/PhysRevB.103.165128} {\bibfield  {journal} {\bibinfo
   {journal} {Phys. Rev. B}\ }\textbf {\bibinfo {volume} {103}},\ \bibinfo
  {pages} {165128} (\bibinfo {year} {2021})}\BibitemShut {NoStop}%
\bibitem [{\citenamefont {Yang}\ \emph {et~al.}(2020)\citenamefont {Yang},
  \citenamefont {Singh}, \citenamefont {Lu}, \citenamefont {Huang},
  \citenamefont {Bahrami}, \citenamefont {Chiu}, \citenamefont {Graf},
  \citenamefont {Huang}, \citenamefont {Wang}, \citenamefont {Lin},
  \citenamefont {Torchinsky}, \citenamefont {Bansil},\ and\ \citenamefont
  {Tafti}}]{yang2020transition}%
  \BibitemOpen
  \bibfield  {author} {\bibinfo {author} {\bibfnamefont {H.-Y.}\ \bibnamefont
  {Yang}}, \bibinfo {author} {\bibfnamefont {B.}~\bibnamefont {Singh}},
  \bibinfo {author} {\bibfnamefont {B.}~\bibnamefont {Lu}}, \bibinfo {author}
  {\bibfnamefont {C.-Y.}\ \bibnamefont {Huang}}, \bibinfo {author}
  {\bibfnamefont {F.}~\bibnamefont {Bahrami}}, \bibinfo {author} {\bibfnamefont
  {W.-C.}\ \bibnamefont {Chiu}}, \bibinfo {author} {\bibfnamefont
  {D.}~\bibnamefont {Graf}}, \bibinfo {author} {\bibfnamefont {S.-M.}\
  \bibnamefont {Huang}}, \bibinfo {author} {\bibfnamefont {B.}~\bibnamefont
  {Wang}}, \bibinfo {author} {\bibfnamefont {H.}~\bibnamefont {Lin}}, \bibinfo
  {author} {\bibfnamefont {D.}~\bibnamefont {Torchinsky}}, \bibinfo {author}
  {\bibfnamefont {A.}~\bibnamefont {Bansil}},\ and\ \bibinfo {author}
  {\bibfnamefont {F.}~\bibnamefont {Tafti}},\ }\bibfield  {title} {\bibinfo
  {title} {{Transition from intrinsic to extrinsic anomalous Hall effect in the
  ferromagnetic Weyl semimetal PrAlGe$_{1-x}$Si$_{x}$}},\ }\href
  {https://doi.org/10.1063/1.5132958} {\bibfield  {journal} {\bibinfo
  {journal} {APL Mater.}\ }\textbf {\bibinfo {volume} {8}},\ \bibinfo {pages}
  {011111} (\bibinfo {year} {2020})}\BibitemShut {NoStop}%
\bibitem [{\citenamefont {Xu}\ \emph {et~al.}(2021{\natexlab{a}})\citenamefont
  {Xu}, \citenamefont {Niu}, \citenamefont {Bai}, \citenamefont {Zhu},
  \citenamefont {Yuan}, \citenamefont {He}, \citenamefont {Yang}, \citenamefont
  {Xia}, \citenamefont {Zhao},\ and\ \citenamefont {Tian}}]{xu2021shubnikov}%
  \BibitemOpen
  \bibfield  {author} {\bibinfo {author} {\bibfnamefont {L.}~\bibnamefont
  {Xu}}, \bibinfo {author} {\bibfnamefont {H.}~\bibnamefont {Niu}}, \bibinfo
  {author} {\bibfnamefont {Y.}~\bibnamefont {Bai}}, \bibinfo {author}
  {\bibfnamefont {H.}~\bibnamefont {Zhu}}, \bibinfo {author} {\bibfnamefont
  {S.}~\bibnamefont {Yuan}}, \bibinfo {author} {\bibfnamefont {X.}~\bibnamefont
  {He}}, \bibinfo {author} {\bibfnamefont {Y.}~\bibnamefont {Yang}}, \bibinfo
  {author} {\bibfnamefont {Z.}~\bibnamefont {Xia}}, \bibinfo {author}
  {\bibfnamefont {L.}~\bibnamefont {Zhao}},\ and\ \bibinfo {author}
  {\bibfnamefont {Z.}~\bibnamefont {Tian}},\ }\bibfield  {title} {\bibinfo
  {title} {{Shubnikov-de Haas Oscillations and Nontrivial Topological State in
  a New Weyl Semimetal Candidate SmAlSi}},\ }\href
  {https://arxiv.org/abs/2107.11957v2} {\bibfield  {journal} {\bibinfo
  {journal} {arXiv:2107.11957}\ } (\bibinfo {year}
  {2021}{\natexlab{a}})}\BibitemShut {NoStop}%
\bibitem [{\citenamefont {Chang}\ \emph {et~al.}(2018)\citenamefont {Chang},
  \citenamefont {Singh}, \citenamefont {Xu}, \citenamefont {Bian},
  \citenamefont {Huang}, \citenamefont {Hsu}, \citenamefont {Belopolski},
  \citenamefont {Alidoust}, \citenamefont {Sanchez}, \citenamefont {Zheng},
  \citenamefont {Lu}, \citenamefont {Zhang}, \citenamefont {Bian},
  \citenamefont {Chang}, \citenamefont {Jeng}, \citenamefont {Bansil},
  \citenamefont {Hsu}, \citenamefont {Jia}, \citenamefont {Neupert},
  \citenamefont {Lin},\ and\ \citenamefont {Hasan}}]{chang2018magnetic}%
  \BibitemOpen
  \bibfield  {author} {\bibinfo {author} {\bibfnamefont {G.}~\bibnamefont
  {Chang}}, \bibinfo {author} {\bibfnamefont {B.}~\bibnamefont {Singh}},
  \bibinfo {author} {\bibfnamefont {S.-Y.}\ \bibnamefont {Xu}}, \bibinfo
  {author} {\bibfnamefont {G.}~\bibnamefont {Bian}}, \bibinfo {author}
  {\bibfnamefont {S.-M.}\ \bibnamefont {Huang}}, \bibinfo {author}
  {\bibfnamefont {C.-H.}\ \bibnamefont {Hsu}}, \bibinfo {author} {\bibfnamefont
  {I.}~\bibnamefont {Belopolski}}, \bibinfo {author} {\bibfnamefont
  {N.}~\bibnamefont {Alidoust}}, \bibinfo {author} {\bibfnamefont {D.~S.}\
  \bibnamefont {Sanchez}}, \bibinfo {author} {\bibfnamefont {H.}~\bibnamefont
  {Zheng}}, \bibinfo {author} {\bibfnamefont {H.}~\bibnamefont {Lu}}, \bibinfo
  {author} {\bibfnamefont {X.}~\bibnamefont {Zhang}}, \bibinfo {author}
  {\bibfnamefont {Y.}~\bibnamefont {Bian}}, \bibinfo {author} {\bibfnamefont
  {T.-R.}\ \bibnamefont {Chang}}, \bibinfo {author} {\bibfnamefont {H.-T.}\
  \bibnamefont {Jeng}}, \bibinfo {author} {\bibfnamefont {A.}~\bibnamefont
  {Bansil}}, \bibinfo {author} {\bibfnamefont {H.}~\bibnamefont {Hsu}},
  \bibinfo {author} {\bibfnamefont {S.}~\bibnamefont {Jia}}, \bibinfo {author}
  {\bibfnamefont {T.}~\bibnamefont {Neupert}}, \bibinfo {author} {\bibfnamefont
  {H.}~\bibnamefont {Lin}},\ and\ \bibinfo {author} {\bibfnamefont {M.~Z.}\
  \bibnamefont {Hasan}},\ }\bibfield  {title} {\bibinfo {title} {{Magnetic and
  noncentrosymmetric Weyl fermion semimetals in the $\mathit{R}\mathrm{AlGe}$
  family of compounds
  ($\mathit{R}=\mathrm{rare}\phantom{\rule{0.28em}{0ex}}\mathrm{earth}$)}},\
  }\href {https://doi.org/10.1103/PhysRevB.97.041104} {\bibfield  {journal}
  {\bibinfo  {journal} {Phys. Rev. B}\ }\textbf {\bibinfo {volume} {97}},\
  \bibinfo {pages} {041104(R)} (\bibinfo {year} {2018})}\BibitemShut {NoStop}%
\bibitem [{\citenamefont {Yang}\ \emph
  {et~al.}(2021{\natexlab{b}})\citenamefont {Yang}, \citenamefont {Singh},
  \citenamefont {Gaudet}, \citenamefont {Lu}, \citenamefont {Huang},
  \citenamefont {Chiu}, \citenamefont {Huang}, \citenamefont {Wang},
  \citenamefont {Bahrami}, \citenamefont {Xu}, \citenamefont {Franklin},
  \citenamefont {Sochnikov}, \citenamefont {Graf}, \citenamefont {Xu},
  \citenamefont {Zhao}, \citenamefont {Hoffman}, \citenamefont {Lin},
  \citenamefont {Torchinsky}, \citenamefont {Broholm}, \citenamefont {Bansil},\
  and\ \citenamefont {Tafti}}]{yang2021noncollinear}%
  \BibitemOpen
  \bibfield  {author} {\bibinfo {author} {\bibfnamefont {H.-Y.}\ \bibnamefont
  {Yang}}, \bibinfo {author} {\bibfnamefont {B.}~\bibnamefont {Singh}},
  \bibinfo {author} {\bibfnamefont {J.}~\bibnamefont {Gaudet}}, \bibinfo
  {author} {\bibfnamefont {B.}~\bibnamefont {Lu}}, \bibinfo {author}
  {\bibfnamefont {C.-Y.}\ \bibnamefont {Huang}}, \bibinfo {author}
  {\bibfnamefont {W.-C.}\ \bibnamefont {Chiu}}, \bibinfo {author}
  {\bibfnamefont {S.-M.}\ \bibnamefont {Huang}}, \bibinfo {author}
  {\bibfnamefont {B.}~\bibnamefont {Wang}}, \bibinfo {author} {\bibfnamefont
  {F.}~\bibnamefont {Bahrami}}, \bibinfo {author} {\bibfnamefont
  {B.}~\bibnamefont {Xu}}, \bibinfo {author} {\bibfnamefont {J.}~\bibnamefont
  {Franklin}}, \bibinfo {author} {\bibfnamefont {I.}~\bibnamefont {Sochnikov}},
  \bibinfo {author} {\bibfnamefont {D.~E.}\ \bibnamefont {Graf}}, \bibinfo
  {author} {\bibfnamefont {G.}~\bibnamefont {Xu}}, \bibinfo {author}
  {\bibfnamefont {Y.}~\bibnamefont {Zhao}}, \bibinfo {author} {\bibfnamefont
  {C.~M.}\ \bibnamefont {Hoffman}}, \bibinfo {author} {\bibfnamefont
  {H.}~\bibnamefont {Lin}}, \bibinfo {author} {\bibfnamefont {D.~H.}\
  \bibnamefont {Torchinsky}}, \bibinfo {author} {\bibfnamefont {C.~L.}\
  \bibnamefont {Broholm}}, \bibinfo {author} {\bibfnamefont {A.}~\bibnamefont
  {Bansil}},\ and\ \bibinfo {author} {\bibfnamefont {F.}~\bibnamefont
  {Tafti}},\ }\bibfield  {title} {\bibinfo {title} {{Noncollinear ferromagnetic
  Weyl semimetal with anisotropic anomalous Hall effect}},\ }\href
  {https://doi.org/10.1103/PhysRevB.103.115143} {\bibfield  {journal} {\bibinfo
   {journal} {Phys. Rev. B}\ }\textbf {\bibinfo {volume} {103}},\ \bibinfo
  {pages} {115143} (\bibinfo {year} {2021}{\natexlab{b}})}\BibitemShut
  {NoStop}%
\bibitem [{\citenamefont {Sanchez}\ \emph {et~al.}(2020)\citenamefont
  {Sanchez}, \citenamefont {Chang}, \citenamefont {Belopolski}, \citenamefont
  {Lu}, \citenamefont {Yin}, \citenamefont {Alidoust}, \citenamefont {Xu},
  \citenamefont {Cochran}, \citenamefont {Zhang}, \citenamefont {Bian} \emph
  {et~al.}}]{sanchez2020observation}%
  \BibitemOpen
  \bibfield  {author} {\bibinfo {author} {\bibfnamefont {D.~S.}\ \bibnamefont
  {Sanchez}}, \bibinfo {author} {\bibfnamefont {G.}~\bibnamefont {Chang}},
  \bibinfo {author} {\bibfnamefont {I.}~\bibnamefont {Belopolski}}, \bibinfo
  {author} {\bibfnamefont {H.}~\bibnamefont {Lu}}, \bibinfo {author}
  {\bibfnamefont {J.-X.}\ \bibnamefont {Yin}}, \bibinfo {author} {\bibfnamefont
  {N.}~\bibnamefont {Alidoust}}, \bibinfo {author} {\bibfnamefont
  {X.}~\bibnamefont {Xu}}, \bibinfo {author} {\bibfnamefont {T.~A.}\
  \bibnamefont {Cochran}}, \bibinfo {author} {\bibfnamefont {X.}~\bibnamefont
  {Zhang}}, \bibinfo {author} {\bibfnamefont {Y.}~\bibnamefont {Bian}}, \emph
  {et~al.},\ }\bibfield  {title} {\bibinfo {title} {{Observation of Weyl
  fermions in a magnetic non-centrosymmetric crystal}},\ }\href
  {https://doi.org/10.1038/s41467-020-16879-1} {\bibfield  {journal} {\bibinfo
  {journal} {Nat. Commun.}\ }\textbf {\bibinfo {volume} {11}},\ \bibinfo
  {pages} {1} (\bibinfo {year} {2020})}\BibitemShut {NoStop}%
\bibitem [{\citenamefont {Destraz}\ \emph {et~al.}(2020)\citenamefont
  {Destraz}, \citenamefont {Das}, \citenamefont {Tsirkin}, \citenamefont {Xu},
  \citenamefont {Neupert}, \citenamefont {Chang}, \citenamefont {Schilling},
  \citenamefont {Grushin}, \citenamefont {Kohlbrecher}, \citenamefont {Keller},
  \citenamefont {Puphal}, \citenamefont {Pomjakushina},\ and\ \citenamefont
  {White}}]{destraz2020magnetism}%
  \BibitemOpen
  \bibfield  {author} {\bibinfo {author} {\bibfnamefont {D.}~\bibnamefont
  {Destraz}}, \bibinfo {author} {\bibfnamefont {L.}~\bibnamefont {Das}},
  \bibinfo {author} {\bibfnamefont {S.~S.}\ \bibnamefont {Tsirkin}}, \bibinfo
  {author} {\bibfnamefont {Y.}~\bibnamefont {Xu}}, \bibinfo {author}
  {\bibfnamefont {T.}~\bibnamefont {Neupert}}, \bibinfo {author} {\bibfnamefont
  {J.}~\bibnamefont {Chang}}, \bibinfo {author} {\bibfnamefont
  {A.}~\bibnamefont {Schilling}}, \bibinfo {author} {\bibfnamefont {A.~G.}\
  \bibnamefont {Grushin}}, \bibinfo {author} {\bibfnamefont {J.}~\bibnamefont
  {Kohlbrecher}}, \bibinfo {author} {\bibfnamefont {L.}~\bibnamefont {Keller}},
  \bibinfo {author} {\bibfnamefont {P.}~\bibnamefont {Puphal}}, \bibinfo
  {author} {\bibfnamefont {E.}~\bibnamefont {Pomjakushina}},\ and\ \bibinfo
  {author} {\bibfnamefont {J.~S.}\ \bibnamefont {White}},\ }\bibfield  {title}
  {\bibinfo {title} {{Magnetism and anomalous transport in the Weyl semimetal
  PrAlGe: possible route to axial gauge fields}},\ }\href
  {https://doi.org/10.1038/s41535-019-0207-7} {\bibfield  {journal} {\bibinfo
  {journal} {npj Quantum Mater.}\ }\textbf {\bibinfo {volume} {5}},\ \bibinfo
  {pages} {5} (\bibinfo {year} {2020})}\BibitemShut {NoStop}%
\bibitem [{\citenamefont {Hodovanets}\ \emph {et~al.}(2018)\citenamefont
  {Hodovanets}, \citenamefont {Eckberg}, \citenamefont {Zavalij}, \citenamefont
  {Kim}, \citenamefont {Lin}, \citenamefont {Zic}, \citenamefont {Campbell},
  \citenamefont {Higgins},\ and\ \citenamefont
  {Paglione}}]{hodovanets2018single}%
  \BibitemOpen
  \bibfield  {author} {\bibinfo {author} {\bibfnamefont {H.}~\bibnamefont
  {Hodovanets}}, \bibinfo {author} {\bibfnamefont {C.~J.}\ \bibnamefont
  {Eckberg}}, \bibinfo {author} {\bibfnamefont {P.~Y.}\ \bibnamefont
  {Zavalij}}, \bibinfo {author} {\bibfnamefont {H.}~\bibnamefont {Kim}},
  \bibinfo {author} {\bibfnamefont {W.-C.}\ \bibnamefont {Lin}}, \bibinfo
  {author} {\bibfnamefont {M.}~\bibnamefont {Zic}}, \bibinfo {author}
  {\bibfnamefont {D.~J.}\ \bibnamefont {Campbell}}, \bibinfo {author}
  {\bibfnamefont {J.~S.}\ \bibnamefont {Higgins}},\ and\ \bibinfo {author}
  {\bibfnamefont {J.}~\bibnamefont {Paglione}},\ }\bibfield  {title} {\bibinfo
  {title} {{Single-crystal investigation of the proposed type-II Weyl semimetal
  CeAlGe}},\ }\href {https://doi.org/10.1103/PhysRevB.98.245132} {\bibfield
  {journal} {\bibinfo  {journal} {Phys. Rev. B}\ }\textbf {\bibinfo {volume}
  {98}},\ \bibinfo {pages} {245132} (\bibinfo {year} {2018})}\BibitemShut
  {NoStop}%
\bibitem [{\citenamefont {Suzuki}\ \emph {et~al.}(2019)\citenamefont {Suzuki},
  \citenamefont {Savary}, \citenamefont {Liu}, \citenamefont {Lynn},
  \citenamefont {Balents},\ and\ \citenamefont
  {Checkelsky}}]{suzuki2019singular}%
  \BibitemOpen
  \bibfield  {author} {\bibinfo {author} {\bibfnamefont {T.}~\bibnamefont
  {Suzuki}}, \bibinfo {author} {\bibfnamefont {L.}~\bibnamefont {Savary}},
  \bibinfo {author} {\bibfnamefont {J.-P.}\ \bibnamefont {Liu}}, \bibinfo
  {author} {\bibfnamefont {J.~W.}\ \bibnamefont {Lynn}}, \bibinfo {author}
  {\bibfnamefont {L.}~\bibnamefont {Balents}},\ and\ \bibinfo {author}
  {\bibfnamefont {J.~G.}\ \bibnamefont {Checkelsky}},\ }\bibfield  {title}
  {\bibinfo {title} {{Singular angular magnetoresistance in a magnetic nodal
  semimetal}},\ }\href {https://doi.org/10.1126/science.aat0348} {\bibfield
  {journal} {\bibinfo  {journal} {Science}\ }\textbf {\bibinfo {volume}
  {365}},\ \bibinfo {pages} {377} (\bibinfo {year} {2019})}\BibitemShut
  {NoStop}%
\bibitem [{\citenamefont {Puphal}\ \emph {et~al.}(2020)\citenamefont {Puphal},
  \citenamefont {Pomjakushin}, \citenamefont {Kanazawa}, \citenamefont
  {Ukleev}, \citenamefont {Gawryluk}, \citenamefont {Ma}, \citenamefont
  {Naamneh}, \citenamefont {Plumb}, \citenamefont {Keller}, \citenamefont
  {Cubitt}, \citenamefont {Pomjakushina},\ and\ \citenamefont
  {White}}]{puphal2020topological}%
  \BibitemOpen
  \bibfield  {author} {\bibinfo {author} {\bibfnamefont {P.}~\bibnamefont
  {Puphal}}, \bibinfo {author} {\bibfnamefont {V.}~\bibnamefont {Pomjakushin}},
  \bibinfo {author} {\bibfnamefont {N.}~\bibnamefont {Kanazawa}}, \bibinfo
  {author} {\bibfnamefont {V.}~\bibnamefont {Ukleev}}, \bibinfo {author}
  {\bibfnamefont {D.~J.}\ \bibnamefont {Gawryluk}}, \bibinfo {author}
  {\bibfnamefont {J.}~\bibnamefont {Ma}}, \bibinfo {author} {\bibfnamefont
  {M.}~\bibnamefont {Naamneh}}, \bibinfo {author} {\bibfnamefont {N.~C.}\
  \bibnamefont {Plumb}}, \bibinfo {author} {\bibfnamefont {L.}~\bibnamefont
  {Keller}}, \bibinfo {author} {\bibfnamefont {R.}~\bibnamefont {Cubitt}},
  \bibinfo {author} {\bibfnamefont {E.}~\bibnamefont {Pomjakushina}},\ and\
  \bibinfo {author} {\bibfnamefont {J.~S.}\ \bibnamefont {White}},\ }\bibfield
  {title} {\bibinfo {title} {{Topological Magnetic Phase in the Candidate Weyl
  Semimetal CeAlGe}},\ }\href {https://doi.org/10.1103/PhysRevLett.124.017202}
  {\bibfield  {journal} {\bibinfo  {journal} {Phys. Rev. Lett.}\ }\textbf
  {\bibinfo {volume} {124}},\ \bibinfo {pages} {017202} (\bibinfo {year}
  {2020})}\BibitemShut {NoStop}%
\bibitem [{\citenamefont {Gaudet}\ \emph {et~al.}()\citenamefont {Gaudet},
  \citenamefont {Yang}, \citenamefont {Baidya}, \citenamefont {Lu},
  \citenamefont {Xu}, \citenamefont {Zhao}, \citenamefont {Rodriguez-Rivera},
  \citenamefont {Hoffmann}, \citenamefont {Graf}, \citenamefont {Torchinsky}
  \emph {et~al.}}]{gaudet2020incommensurate}%
  \BibitemOpen
  \bibfield  {author} {\bibinfo {author} {\bibfnamefont {J.}~\bibnamefont
  {Gaudet}}, \bibinfo {author} {\bibfnamefont {H.-Y.}\ \bibnamefont {Yang}},
  \bibinfo {author} {\bibfnamefont {S.}~\bibnamefont {Baidya}}, \bibinfo
  {author} {\bibfnamefont {B.}~\bibnamefont {Lu}}, \bibinfo {author}
  {\bibfnamefont {G.}~\bibnamefont {Xu}}, \bibinfo {author} {\bibfnamefont
  {Y.}~\bibnamefont {Zhao}}, \bibinfo {author} {\bibfnamefont {J.~A.}\
  \bibnamefont {Rodriguez-Rivera}}, \bibinfo {author} {\bibfnamefont {C.~M.}\
  \bibnamefont {Hoffmann}}, \bibinfo {author} {\bibfnamefont {D.~E.}\
  \bibnamefont {Graf}}, \bibinfo {author} {\bibfnamefont {D.~H.}\ \bibnamefont
  {Torchinsky}}, \emph {et~al.},\ }\bibfield  {title} {\bibinfo {title}
  {{Weyl-mediated helical magnetism in NdAlSi}},\ }\href@noop {} {\bibinfo
  {journal} {Nat. Mater.}\ }\BibitemShut {NoStop}%
\bibitem [{\citenamefont {Flandorfer}\ \emph {et~al.}(1998)\citenamefont
  {Flandorfer}, \citenamefont {Kaczorowski}, \citenamefont {Gr{\"o}bner},
  \citenamefont {Rogl}, \citenamefont {Wouters}, \citenamefont {Godart},\ and\
  \citenamefont {Kostikas}}]{flandorfer1998systems}%
  \BibitemOpen
\bibfield  {journal} {  }\bibfield  {author} {\bibinfo {author} {\bibfnamefont
  {H.}~\bibnamefont {Flandorfer}}, \bibinfo {author} {\bibfnamefont
  {D.}~\bibnamefont {Kaczorowski}}, \bibinfo {author} {\bibfnamefont
  {J.}~\bibnamefont {Gr{\"o}bner}}, \bibinfo {author} {\bibfnamefont
  {P.}~\bibnamefont {Rogl}}, \bibinfo {author} {\bibfnamefont {R.}~\bibnamefont
  {Wouters}}, \bibinfo {author} {\bibfnamefont {C.}~\bibnamefont {Godart}},\
  and\ \bibinfo {author} {\bibfnamefont {A.}~\bibnamefont {Kostikas}},\
  }\bibfield  {title} {\bibinfo {title} {{The systems Ce--Al--(Si, Ge): phase
  equilibria and physical properties}},\ }\href
  {https://doi.org/10.1006/jssc.1997.7660} {\bibfield  {journal} {\bibinfo
  {journal} {J. Solid State Chem.}\ }\textbf {\bibinfo {volume} {137}},\
  \bibinfo {pages} {191} (\bibinfo {year} {1998})}\BibitemShut {NoStop}%
\bibitem [{\citenamefont {Pukas}\ \emph {et~al.}(2004)\citenamefont {Pukas},
  \citenamefont {Lutsyshyn}, \citenamefont {Manyako},\ and\ \citenamefont
  {Gladyshevskii}}]{pukas2004crystal}%
  \BibitemOpen
  \bibfield  {author} {\bibinfo {author} {\bibfnamefont {S.}~\bibnamefont
  {Pukas}}, \bibinfo {author} {\bibfnamefont {Y.}~\bibnamefont {Lutsyshyn}},
  \bibinfo {author} {\bibfnamefont {M.}~\bibnamefont {Manyako}},\ and\ \bibinfo
  {author} {\bibfnamefont {E.}~\bibnamefont {Gladyshevskii}},\ }\bibfield
  {title} {\bibinfo {title} {{Crystal structures of the RAlSi and RAlGe
  compounds}},\ }\href {https://doi.org/10.1016/j.jallcom.2003.08.031}
  {\bibfield  {journal} {\bibinfo  {journal} {J. Alloys Compd.}\ }\textbf
  {\bibinfo {volume} {367}},\ \bibinfo {pages} {162} (\bibinfo {year}
  {2004})}\BibitemShut {NoStop}%
\bibitem [{\citenamefont {Bobev}\ \emph {et~al.}(2005)\citenamefont {Bobev},
  \citenamefont {Tobash}, \citenamefont {Fritsch}, \citenamefont {Thompson},
  \citenamefont {Hundley}, \citenamefont {Sarrao},\ and\ \citenamefont
  {Fisk}}]{bobev2005ternary}%
  \BibitemOpen
  \bibfield  {author} {\bibinfo {author} {\bibfnamefont {S.}~\bibnamefont
  {Bobev}}, \bibinfo {author} {\bibfnamefont {P.~H.}\ \bibnamefont {Tobash}},
  \bibinfo {author} {\bibfnamefont {V.}~\bibnamefont {Fritsch}}, \bibinfo
  {author} {\bibfnamefont {J.~D.}\ \bibnamefont {Thompson}}, \bibinfo {author}
  {\bibfnamefont {M.~F.}\ \bibnamefont {Hundley}}, \bibinfo {author}
  {\bibfnamefont {J.~L.}\ \bibnamefont {Sarrao}},\ and\ \bibinfo {author}
  {\bibfnamefont {Z.}~\bibnamefont {Fisk}},\ }\bibfield  {title} {\bibinfo
  {title} {{Ternary rare-earth alumo-silicides—single-crystal growth from Al
  flux, structural and physical properties}},\ }\href
  {https://doi.org/10.1016/j.jssc.2005.04.021} {\bibfield  {journal} {\bibinfo
  {journal} {J. Solid State Chem.}\ }\textbf {\bibinfo {volume} {178}},\
  \bibinfo {pages} {2091} (\bibinfo {year} {2005})}\BibitemShut {NoStop}%
\bibitem [{\citenamefont {Sharma}\ \emph {et~al.}(2007)\citenamefont {Sharma},
  \citenamefont {Bobev},\ and\ \citenamefont
  {Sarrao}}]{sharma2007oscillations}%
  \BibitemOpen
  \bibfield  {author} {\bibinfo {author} {\bibfnamefont {A.~L.}\ \bibnamefont
  {Sharma}}, \bibinfo {author} {\bibfnamefont {S.}~\bibnamefont {Bobev}},\ and\
  \bibinfo {author} {\bibfnamefont {J.}~\bibnamefont {Sarrao}},\ }\bibfield
  {title} {\bibinfo {title} {{Oscillations in magnetocaloric effect and
  magnetic properties of RE$_{2}$Al$_{3}$Si$_{2}$ (for RE=Dy, Ho and Er) and
  REAlSi (for RE=Ce and Pr)}},\ }\href
  {https://doi.org/10.1016/j.jmmm.2006.11.125} {\bibfield  {journal} {\bibinfo
  {journal} {J. Magn. Magn. Mater.}\ }\textbf {\bibinfo {volume} {312}},\
  \bibinfo {pages} {400} (\bibinfo {year} {2007})}\BibitemShut {NoStop}%
\bibitem [{\citenamefont {Xu}\ \emph {et~al.}(2021{\natexlab{b}})\citenamefont
  {Xu}, \citenamefont {Franklin}, \citenamefont {Jayakody}, \citenamefont
  {Yang}, \citenamefont {Tafti},\ and\ \citenamefont
  {Sochnikov}}]{xu2021picoscale}%
  \BibitemOpen
  \bibfield  {author} {\bibinfo {author} {\bibfnamefont {B.}~\bibnamefont
  {Xu}}, \bibinfo {author} {\bibfnamefont {J.}~\bibnamefont {Franklin}},
  \bibinfo {author} {\bibfnamefont {A.}~\bibnamefont {Jayakody}}, \bibinfo
  {author} {\bibfnamefont {H.-Y.}\ \bibnamefont {Yang}}, \bibinfo {author}
  {\bibfnamefont {F.}~\bibnamefont {Tafti}},\ and\ \bibinfo {author}
  {\bibfnamefont {I.}~\bibnamefont {Sochnikov}},\ }\bibfield  {title} {\bibinfo
  {title} {{Picoscale Magnetoelasticity Governs Heterogeneous Magnetic Domains
  in a Noncentrosymmetric Ferromagnetic Weyl Semimetal}},\ }\href
  {https://doi.org/https://doi.org/10.1002/qute.202000101} {\bibfield
  {journal} {\bibinfo  {journal} {Adv. Quantum Tech.}\ }\textbf {\bibinfo
  {volume} {4}},\ \bibinfo {pages} {2000101} (\bibinfo {year}
  {2021}{\natexlab{b}})}\BibitemShut {NoStop}%
\bibitem [{\citenamefont {Huang}\ \emph {et~al.}(2021)\citenamefont {Huang},
  \citenamefont {Lane}, \citenamefont {Yarotski}, \citenamefont {Taylor},\ and\
  \citenamefont {Zhu}}]{huang2021topological}%
  \BibitemOpen
  \bibfield  {author} {\bibinfo {author} {\bibfnamefont {Z.}~\bibnamefont
  {Huang}}, \bibinfo {author} {\bibfnamefont {C.}~\bibnamefont {Lane}},
  \bibinfo {author} {\bibfnamefont {D.}~\bibnamefont {Yarotski}}, \bibinfo
  {author} {\bibfnamefont {A.}~\bibnamefont {Taylor}},\ and\ \bibinfo {author}
  {\bibfnamefont {J.-X.}\ \bibnamefont {Zhu}},\ }\bibfield  {title} {\bibinfo
  {title} {{Topological superconducting domain walls in magnetic Weyl
  semimetals}},\ }\href {https://arxiv.org/abs/2106.02215} {\bibfield
  {journal} {\bibinfo  {journal} {arXiv:2106.02215}\ } (\bibinfo {year}
  {2021})}\BibitemShut {NoStop}%
\bibitem [{\citenamefont {Sun}\ \emph {et~al.}(2021)\citenamefont {Sun},
  \citenamefont {Lee}, \citenamefont {Yang}, \citenamefont {Torchinsky},
  \citenamefont {Tafti},\ and\ \citenamefont {Orenstein}}]{sun2021mapping}%
  \BibitemOpen
  \bibfield  {author} {\bibinfo {author} {\bibfnamefont {Y.}~\bibnamefont
  {Sun}}, \bibinfo {author} {\bibfnamefont {C.}~\bibnamefont {Lee}}, \bibinfo
  {author} {\bibfnamefont {H.-Y.}\ \bibnamefont {Yang}}, \bibinfo {author}
  {\bibfnamefont {D.~H.}\ \bibnamefont {Torchinsky}}, \bibinfo {author}
  {\bibfnamefont {F.}~\bibnamefont {Tafti}},\ and\ \bibinfo {author}
  {\bibfnamefont {J.}~\bibnamefont {Orenstein}},\ }\bibfield  {title} {\bibinfo
  {title} {{Mapping Domain Wall Topology in the Magnetic Weyl Semimetal
  CeAlSi}},\ }\href {https://arxiv.org/abs/2104.07706} {\bibfield  {journal}
  {\bibinfo  {journal} {arXiv:2104.07706}\ } (\bibinfo {year}
  {2021})}\BibitemShut {NoStop}%
\bibitem [{\citenamefont {dos Reis}\ \emph {et~al.}(2016)\citenamefont {dos
  Reis}, \citenamefont {Wu}, \citenamefont {Sun}, \citenamefont {Ajeesh},
  \citenamefont {Shekhar}, \citenamefont {Schmidt}, \citenamefont {Felser},
  \citenamefont {Yan},\ and\ \citenamefont {Nicklas}}]{dos2016pressure}%
  \BibitemOpen
  \bibfield  {author} {\bibinfo {author} {\bibfnamefont {R.~D.}\ \bibnamefont
  {dos Reis}}, \bibinfo {author} {\bibfnamefont {S.~C.}\ \bibnamefont {Wu}},
  \bibinfo {author} {\bibfnamefont {Y.}~\bibnamefont {Sun}}, \bibinfo {author}
  {\bibfnamefont {M.~O.}\ \bibnamefont {Ajeesh}}, \bibinfo {author}
  {\bibfnamefont {C.}~\bibnamefont {Shekhar}}, \bibinfo {author} {\bibfnamefont
  {M.}~\bibnamefont {Schmidt}}, \bibinfo {author} {\bibfnamefont
  {C.}~\bibnamefont {Felser}}, \bibinfo {author} {\bibfnamefont
  {B.}~\bibnamefont {Yan}},\ and\ \bibinfo {author} {\bibfnamefont
  {M.}~\bibnamefont {Nicklas}},\ }\bibfield  {title} {\bibinfo {title}
  {{Pressure tuning the Fermi surface topology of the Weyl semimetal NbP}},\
  }\href {https://doi.org/10.1103/PhysRevB.93.205102} {\bibfield  {journal}
  {\bibinfo  {journal} {Phys. Rev. B}\ }\textbf {\bibinfo {volume} {93}},\
  \bibinfo {pages} {205102} (\bibinfo {year} {2016})}\BibitemShut {NoStop}%
\bibitem [{\citenamefont {Liang}\ \emph {et~al.}(2017)\citenamefont {Liang},
  \citenamefont {Kushwaha}, \citenamefont {Kim}, \citenamefont {Gibson},
  \citenamefont {Lin}, \citenamefont {Kioussis}, \citenamefont {Cava},\ and\
  \citenamefont {Ong}}]{liang2017pressure}%
  \BibitemOpen
  \bibfield  {author} {\bibinfo {author} {\bibfnamefont {T.}~\bibnamefont
  {Liang}}, \bibinfo {author} {\bibfnamefont {S.}~\bibnamefont {Kushwaha}},
  \bibinfo {author} {\bibfnamefont {J.}~\bibnamefont {Kim}}, \bibinfo {author}
  {\bibfnamefont {Q.}~\bibnamefont {Gibson}}, \bibinfo {author} {\bibfnamefont
  {J.}~\bibnamefont {Lin}}, \bibinfo {author} {\bibfnamefont {N.}~\bibnamefont
  {Kioussis}}, \bibinfo {author} {\bibfnamefont {R.~J.}\ \bibnamefont {Cava}},\
  and\ \bibinfo {author} {\bibfnamefont {N.~P.}\ \bibnamefont {Ong}},\
  }\bibfield  {title} {\bibinfo {title} {{A pressure-induced topological phase
  with large Berry curvature in Pb$_{1-x}$Sn$_{x}$Te}},\ }\href
  {https://doi.org/10.1126/sciadv.1602510} {\bibfield  {journal} {\bibinfo
  {journal} {Sci. Adv.}\ }\textbf {\bibinfo {volume} {3}},\ \bibinfo {pages}
  {e1602510} (\bibinfo {year} {2017})}\BibitemShut {NoStop}%
\bibitem [{\citenamefont {Hirayama}\ \emph {et~al.}(2015)\citenamefont
  {Hirayama}, \citenamefont {Okugawa}, \citenamefont {Ishibashi}, \citenamefont
  {Murakami},\ and\ \citenamefont {Miyake}}]{hirayama2015weyl}%
  \BibitemOpen
  \bibfield  {author} {\bibinfo {author} {\bibfnamefont {M.}~\bibnamefont
  {Hirayama}}, \bibinfo {author} {\bibfnamefont {R.}~\bibnamefont {Okugawa}},
  \bibinfo {author} {\bibfnamefont {S.}~\bibnamefont {Ishibashi}}, \bibinfo
  {author} {\bibfnamefont {S.}~\bibnamefont {Murakami}},\ and\ \bibinfo
  {author} {\bibfnamefont {T.}~\bibnamefont {Miyake}},\ }\bibfield  {title}
  {\bibinfo {title} {{Weyl Node and Spin Texture in Trigonal Tellurium and
  Selenium}},\ }\href {https://doi.org/10.1103/PhysRevLett.114.206401}
  {\bibfield  {journal} {\bibinfo  {journal} {Phys. Rev. Lett.}\ }\textbf
  {\bibinfo {volume} {114}},\ \bibinfo {pages} {206401} (\bibinfo {year}
  {2015})}\BibitemShut {NoStop}%
\bibitem [{\citenamefont {Rodriguez}\ \emph {et~al.}(2020)\citenamefont
  {Rodriguez}, \citenamefont {Tsirlin}, \citenamefont {Biesner}, \citenamefont
  {Ueno}, \citenamefont {Takahashi}, \citenamefont {Kobayashi}, \citenamefont
  {Dressel},\ and\ \citenamefont {Uykur}}]{rodriguez2020two}%
  \BibitemOpen
  \bibfield  {author} {\bibinfo {author} {\bibfnamefont {D.}~\bibnamefont
  {Rodriguez}}, \bibinfo {author} {\bibfnamefont {A.~A.}\ \bibnamefont
  {Tsirlin}}, \bibinfo {author} {\bibfnamefont {T.}~\bibnamefont {Biesner}},
  \bibinfo {author} {\bibfnamefont {T.}~\bibnamefont {Ueno}}, \bibinfo {author}
  {\bibfnamefont {T.}~\bibnamefont {Takahashi}}, \bibinfo {author}
  {\bibfnamefont {K.}~\bibnamefont {Kobayashi}}, \bibinfo {author}
  {\bibfnamefont {M.}~\bibnamefont {Dressel}},\ and\ \bibinfo {author}
  {\bibfnamefont {E.}~\bibnamefont {Uykur}},\ }\bibfield  {title} {\bibinfo
  {title} {{Two Linear Regimes in Optical Conductivity of a Type-I Weyl
  Semimetal: The Case of Elemental Tellurium}},\ }\href
  {https://doi.org/10.1103/PhysRevLett.124.136402} {\bibfield  {journal}
  {\bibinfo  {journal} {Phys. Rev. Lett.}\ }\textbf {\bibinfo {volume} {124}},\
  \bibinfo {pages} {136402} (\bibinfo {year} {2020})}\BibitemShut {NoStop}%
\bibitem [{\citenamefont {Ma}\ \emph {et~al.}(2015)\citenamefont {Ma},
  \citenamefont {Cui}, \citenamefont {Ueda}, \citenamefont {Tang},
  \citenamefont {Chen}, \citenamefont {Tamura}, \citenamefont {Wu},
  \citenamefont {Fujioka}, \citenamefont {Tokura},\ and\ \citenamefont
  {Shen}}]{ma2015mobile}%
  \BibitemOpen
  \bibfield  {author} {\bibinfo {author} {\bibfnamefont {E.~Y.}\ \bibnamefont
  {Ma}}, \bibinfo {author} {\bibfnamefont {Y.-T.}\ \bibnamefont {Cui}},
  \bibinfo {author} {\bibfnamefont {K.}~\bibnamefont {Ueda}}, \bibinfo {author}
  {\bibfnamefont {S.}~\bibnamefont {Tang}}, \bibinfo {author} {\bibfnamefont
  {K.}~\bibnamefont {Chen}}, \bibinfo {author} {\bibfnamefont {N.}~\bibnamefont
  {Tamura}}, \bibinfo {author} {\bibfnamefont {P.~M.}\ \bibnamefont {Wu}},
  \bibinfo {author} {\bibfnamefont {J.}~\bibnamefont {Fujioka}}, \bibinfo
  {author} {\bibfnamefont {Y.}~\bibnamefont {Tokura}},\ and\ \bibinfo {author}
  {\bibfnamefont {Z.-X.}\ \bibnamefont {Shen}},\ }\bibfield  {title} {\bibinfo
  {title} {{Mobile metallic domain walls in an all-in-all-out magnetic
  insulator}},\ }\href {https://doi.org/10.1126/science.aac8289} {\bibfield
  {journal} {\bibinfo  {journal} {Science}\ }\textbf {\bibinfo {volume}
  {350}},\ \bibinfo {pages} {538} (\bibinfo {year} {2015})}\BibitemShut
  {NoStop}%
\bibitem [{\citenamefont {Ueda}\ \emph {et~al.}(2018)\citenamefont {Ueda},
  \citenamefont {Kaneko}, \citenamefont {Ishizuka}, \citenamefont {Fujioka},
  \citenamefont {Nagaosa},\ and\ \citenamefont {Tokura}}]{ueda2018spontaneous}%
  \BibitemOpen
  \bibfield  {author} {\bibinfo {author} {\bibfnamefont {K.}~\bibnamefont
  {Ueda}}, \bibinfo {author} {\bibfnamefont {R.}~\bibnamefont {Kaneko}},
  \bibinfo {author} {\bibfnamefont {H.}~\bibnamefont {Ishizuka}}, \bibinfo
  {author} {\bibfnamefont {J.}~\bibnamefont {Fujioka}}, \bibinfo {author}
  {\bibfnamefont {N.}~\bibnamefont {Nagaosa}},\ and\ \bibinfo {author}
  {\bibfnamefont {Y.}~\bibnamefont {Tokura}},\ }\bibfield  {title} {\bibinfo
  {title} {{Spontaneous Hall effect in the Weyl semimetal candidate of all-in
  all-out pyrochlore iridate}},\ }\href
  {https://doi.org/10.1038/s41467-018-05530-9} {\bibfield  {journal} {\bibinfo
  {journal} {Nat. Commun.}\ }\textbf {\bibinfo {volume} {9}},\ \bibinfo {pages}
  {1} (\bibinfo {year} {2018})}\BibitemShut {NoStop}%
\bibitem [{\citenamefont {Yamaji}\ and\ \citenamefont
  {Imada}(2014)}]{yamaji2014metallic}%
  \BibitemOpen
  \bibfield  {author} {\bibinfo {author} {\bibfnamefont {Y.}~\bibnamefont
  {Yamaji}}\ and\ \bibinfo {author} {\bibfnamefont {M.}~\bibnamefont {Imada}},\
  }\bibfield  {title} {\bibinfo {title} {{Metallic Interface Emerging at
  Magnetic Domain Wall of Antiferromagnetic Insulator: Fate of Extinct Weyl
  Electrons}},\ }\href {https://doi.org/10.1103/PhysRevX.4.021035} {\bibfield
  {journal} {\bibinfo  {journal} {Phys. Rev. X}\ }\textbf {\bibinfo {volume}
  {4}},\ \bibinfo {pages} {021035} (\bibinfo {year} {2014})}\BibitemShut
  {NoStop}%
\bibitem [{\citenamefont {Lovett}(2018)}]{lovett2018tensor}%
  \BibitemOpen
  \bibfield  {author} {\bibinfo {author} {\bibfnamefont {D.~R.}\ \bibnamefont
  {Lovett}},\ }\href@noop {} {\emph {\bibinfo {title} {Tensor properties of
  crystals}}}\ (\bibinfo  {publisher} {CRC Press},\ \bibinfo {year}
  {2018})\BibitemShut {NoStop}%
\bibitem [{\citenamefont {Shoenberg}(2009)}]{shoenberg2009magnetic}%
  \BibitemOpen
  \bibfield  {author} {\bibinfo {author} {\bibfnamefont {D.}~\bibnamefont
  {Shoenberg}},\ }\href@noop {} {\emph {\bibinfo {title} {Magnetic oscillations
  in metals}}}\ (\bibinfo  {publisher} {Cambridge University Press},\ \bibinfo
  {year} {2009})\BibitemShut {NoStop}%
\bibitem [{\citenamefont {Honold}\ \emph {et~al.}(1997)\citenamefont {Honold},
  \citenamefont {Harrison}, \citenamefont {Singleton}, \citenamefont {Yaguchi},
  \citenamefont {Mielke}, \citenamefont {Rickel}, \citenamefont {Deckers},
  \citenamefont {Reinders}, \citenamefont {Herlach}, \citenamefont {Kurmoo},\
  and\ \citenamefont {Day}}]{honold1997importance}%
  \BibitemOpen
  \bibfield  {author} {\bibinfo {author} {\bibfnamefont {M.~M.}\ \bibnamefont
  {Honold}}, \bibinfo {author} {\bibfnamefont {N.}~\bibnamefont {Harrison}},
  \bibinfo {author} {\bibfnamefont {J.}~\bibnamefont {Singleton}}, \bibinfo
  {author} {\bibfnamefont {H.}~\bibnamefont {Yaguchi}}, \bibinfo {author}
  {\bibfnamefont {C.}~\bibnamefont {Mielke}}, \bibinfo {author} {\bibfnamefont
  {D.}~\bibnamefont {Rickel}}, \bibinfo {author} {\bibfnamefont
  {I.}~\bibnamefont {Deckers}}, \bibinfo {author} {\bibfnamefont {P.~H.~P.}\
  \bibnamefont {Reinders}}, \bibinfo {author} {\bibfnamefont {F.}~\bibnamefont
  {Herlach}}, \bibinfo {author} {\bibfnamefont {M.}~\bibnamefont {Kurmoo}},\
  and\ \bibinfo {author} {\bibfnamefont {P.}~\bibnamefont {Day}},\ }\bibfield
  {title} {\bibinfo {title} {{The importance of edge states in the quantum Hall
  regime of the organic conductor}},\ }\href
  {https://doi.org/10.1088/0953-8984/9/39/001} {\bibfield  {journal} {\bibinfo
  {journal} {J. Condens. Matter Phys.}\ }\textbf {\bibinfo {volume} {9}},\
  \bibinfo {pages} {L533} (\bibinfo {year} {1997})}\BibitemShut {NoStop}%
\bibitem [{\citenamefont {Wu}\ \emph {et~al.}(2019)\citenamefont {Wu},
  \citenamefont {Guo}, \citenamefont {Smidman}, \citenamefont {Zhang},
  \citenamefont {Chen}, \citenamefont {Singleton},\ and\ \citenamefont
  {Yuan}}]{wu2019anomalous}%
  \BibitemOpen
  \bibfield  {author} {\bibinfo {author} {\bibfnamefont {F.}~\bibnamefont
  {Wu}}, \bibinfo {author} {\bibfnamefont {C.}~\bibnamefont {Guo}}, \bibinfo
  {author} {\bibfnamefont {M.}~\bibnamefont {Smidman}}, \bibinfo {author}
  {\bibfnamefont {J.}~\bibnamefont {Zhang}}, \bibinfo {author} {\bibfnamefont
  {Y.}~\bibnamefont {Chen}}, \bibinfo {author} {\bibfnamefont {J.}~\bibnamefont
  {Singleton}},\ and\ \bibinfo {author} {\bibfnamefont {H.}~\bibnamefont
  {Yuan}},\ }\bibfield  {title} {\bibinfo {title} {{Anomalous quantum
  oscillations and evidence for a non-trivial Berry phase in SmSb}},\ }\href
  {https://doi.org/10.1038/s41535-019-0161-4} {\bibfield  {journal} {\bibinfo
  {journal} {npj Quantum Mater.}\ }\textbf {\bibinfo {volume} {4}},\ \bibinfo
  {pages} {1} (\bibinfo {year} {2019})}\BibitemShut {NoStop}%
\bibitem [{\citenamefont {Wang}\ \emph {et~al.}(2016)\citenamefont {Wang},
  \citenamefont {Lu},\ and\ \citenamefont {Shen}}]{wang2016anomalous}%
  \BibitemOpen
  \bibfield  {author} {\bibinfo {author} {\bibfnamefont {C.~M.}\ \bibnamefont
  {Wang}}, \bibinfo {author} {\bibfnamefont {H.-Z.}\ \bibnamefont {Lu}},\ and\
  \bibinfo {author} {\bibfnamefont {S.-Q.}\ \bibnamefont {Shen}},\ }\bibfield
  {title} {\bibinfo {title} {{Anomalous Phase Shift of Quantum Oscillations in
  3D Topological Semimetals}},\ }\href
  {https://doi.org/10.1103/PhysRevLett.117.077201} {\bibfield  {journal}
  {\bibinfo  {journal} {Phys. Rev. Lett.}\ }\textbf {\bibinfo {volume} {117}},\
  \bibinfo {pages} {077201} (\bibinfo {year} {2016})}\BibitemShut {NoStop}%
\end{thebibliography}%

\end{document}